\documentclass[preprint,showpacs,preprintnumbers,amsmath,amssymb]{revtex4}
\usepackage{booktabs}
\usepackage{mathrsfs}
\usepackage{epsfig}
\usepackage{graphicx}
\usepackage{dcolumn}
\usepackage{bm}
\usepackage{amsmath}
\usepackage{slashed}       

\let\jnfont=\rm
\def\NPB#1,{{\jnfont Nucl.\ Phys.\ B }{\bf #1},}
\def\PLB#1,{{\jnfont Phys.\ Lett.\ B }{\bf #1},}
\def\EPJC#1,{{\jnfont Eur.\ Phys.\ Jour.\ C }{\bf #1},}
\def\PRD#1,{{\jnfont Phys.\ Rev.\ D }{\bf #1},}
\def\PRL#1,{{\jnfont Phys.\ Rev.\ Lett.\ }{\bf #1},}
\def\MPLA#1,{{\jnfont Mod.\ Phys.\ Lett.\ A }{\bf #1},}
\def\JPG#1,{{\jnfont J.\ Phys.\ G}{\bf #1},}
\def\CTP#1,{{\jnfont Commun.\ Theor.\ Phys.\ }{\bf #1},}
\def\ZPC#1,{{\jnfont Z.\ Phys.\ C }{\bf #1},}
\def\JHEP#1,{{\jnfont JHEP \ }{\bf #1},}
\def\lsim{\raise0.3ex\hbox{$<$\kern-0.75em\raise-1.1ex\hbox{$\sim$}}}
\def\gsim{\raise0.3ex\hbox{$>$\kern-0.75em\raise-1.1ex\hbox{$\sim$}}}

\begin{document}

\title{A SM-like Higgs near 125 GeV in low energy SUSY: a comparative study for MSSM and NMSSM}

\author{Junjie Cao$^{1,2}$, Zhaoxia Heng$^1$, Jin Min Yang$^3$, Yanming Zhang$^1$, Jingya Zhu$^3$, }

\affiliation{
  $^1$  Department of Physics,
        Henan Normal University, Xinxiang 453007, China \\
  $^2$ Center for High Energy Physics, Peking University,
       Beijing 100871, China \\
  $^3$ State Key Laboratory of Theoretical Physics,
      Institute of Theoretical Physics, Academia Sinica, Beijing 100190, China
      \vspace{1cm}}

\begin{abstract}
Motivated by the recent LHC hints of a Higgs boson around 125 GeV,
we assume a SM-like Higgs with the mass 123-127 GeV and study its
implication in low energy SUSY by comparing the MSSM and NMSSM.
We consider various experimental constraints at $2\sigma$ level
(including the muon $g-2$ and the dark matter relic density) and
perform a comprehensive scan over the parameter space of each model.
Then in the parameter space which is allowed by current experimental constraints
and also predicts a SM-like Higgs in 123-127 GeV, we examine the properties
of the sensitive parameters (like the top squark mass and the trilinear
coupling $A_t$) and calculate the rates of the di-photon signal and
the $VV^*$ ($V=W,Z$) signals at the LHC. Our typical findings are: (i)  In the MSSM
the top squark and $A_t$ must be large and thus incur some fine-tuning,
which can be much ameliorated in the NMSSM; (ii) In the MSSM a light stau
is needed to enhance the di-photon rate of the SM-like Higgs to exceed its SM
prediction, while in the NMSSM the di-photon rate can be readily enhanced
in several ways;
(iii) In the MSSM the signal rates of $pp\to h\to VV^*$ at the LHC 
are never enhanced compared with their SM predictions, 
while in the NMSSM they may get enhanced significantly;
(iv) A large part of the parameter space so far survived will be soon covered by the expected
XENON100(2012) sensitivity (especially for the NMSSM).
\end{abstract}

\maketitle

\section{Introduction}
As the only missing particle in the Standard Model (SM), the Higgs boson is being intensively
hunted at the LHC. Recently, both the ATLAS and CMS experiments have revealed hints of a
Higgs particle around $125 {\rm GeV}$ \cite{ATLAS,CMS}. While such a Higgs mass
can be well accommodated in the SM and the reported
signal rates in several channels are also in agreement with the SM expectations after
taking into account the large experimental uncertainty \cite{Carmi,Espinosa}
(albeit the central value of the observed di-photon rate is somewhat above the SM prediction),
the low energy
supersymmetry (SUSY) seems to be a better framework to account for such a Higgs.
In low energy SUSY the SM-like Higgs mass is theoretically restricted in a narrow range
and its di-photon rate at the LHC may exceed the SM prediction
\cite{diphoton1,diphoton2,diphoton3}, both of which are welcomed by the LHC results.

However, as the most popular low energy SUSY model, the minimal supersymmetric standard model
(MSSM) \cite{Haber,Djouadi} may have some tension to accommodate such a 125 GeV Higgs.
As is well known, in the MSSM the SM-like Higgs mass is upper bounded by
$M_Z\cos2\beta$ at tree-level and to get a Higgs around $125 {\rm GeV}$ we need sizable
top/stop loop contributions, which depend quartically on the top quark mass and
logarithmically on the stop masses \cite{Carena:1995bx}.
This will, on the one hand, impose rather tight constraint on the MSSM, and on the other hand,
incur some fine-tuning \cite{tuning}.
Such a problem can be alleviated in the so-called next-to-minimal supersymmetric standard model
(NMSSM) \cite{NMSSM1,NMSSM2}, which is the simplest singlet extension of the MSSM with a scale
invariant superpotential.
In the NMSSM, due to the introduction of some new couplings in the superpotential,
the SM-like Higgs mass gets additional contribution at tree-level and also may be
further pushed up by the mixing effect in diagonalizing the mass matrix of the CP-even
Higgs fields \cite{king,Kang:2012tn}.
As a result, a SM-like Higgs around $125 {\rm GeV}$ does not entail large loop contributions,
which may thus ameliorate the fine-tuning problem \cite{tuning2}.

In this work, motivated by the recent LHC results, we assume a SM-like Higgs boson in
$123-127~ {\rm GeV}$ and study its implication in the MSSM and NMSSM.
Different from recent studies in this direction
\cite{diphoton2,125Higgs,Carena:2011aa,king,Kang:2012tn},
we scan the model parameters by considering various experimental constraints
and perform a comparative study for the MSSM and NMSSM.
We will investigate the features of the allowed parameter space in each model
and particularly we pay more attention to the space of the NMSSM which may be
distinct from the MSSM. Noting the LHC experiments utilize the channels
$ p p \to h \to \gamma \gamma$,
$ p p \to h \to Z Z^\ast \to 4 l $ and $ p p \to h \to W W^\ast \to 2 l\ 2 \nu$ in
 searching for the Higgs boson \cite{ATLAS,CMS},
we also study their normalized production rates defined as
 \begin{eqnarray}
R_{\gamma \gamma}&\equiv & \sigma_{SUSY} ( p p \to h \to
\gamma \gamma)/\sigma_{SM} ( p p \to h \to \gamma \gamma )\nonumber\\
&\equiv & C_{hgg}^2 C_{h\gamma\gamma}^2 \times
\Gamma_{tot}(h_{SM})/\Gamma_{tot}(h), \label{definition1}\\
R_{V V}&\equiv & \sigma_{SUSY} ( p p \to h \to
VV^\ast)/\sigma_{SM} ( p p \to h \to VV^\ast)\nonumber\\
&\equiv & C_{hgg}^2 C_{hVV}^2 \times
\Gamma_{tot}(h_{SM})/\Gamma_{tot}(h), \label{definition2}
\end{eqnarray}
where $V= W, Z$, and $C_{hgg}$, $C_{h\gamma\gamma}$ and $C_{hVV}$ are respectively the rescaled
couplings of the Higgs to gluons, photons and weak gauge bosons by their SM values.
We are more interested in the case with $R_{\gamma \gamma} > 1$ because it is favored
by current ATLAS and CMS results \cite{ATLAS,CMS,Carmi}. As will be shown below,
this case usually predicts a slepton or a chargino lighter than $250~ {\rm GeV}$.

This paper is organized as follows. In Sec. II, we recapitulate the characters of the
Higgs mass in the models for better understanding our numerical results. In Sec. III,
we perform a comprehensive scan over the parameter space of each model
by imposing current experimental constraints and also by requiring
a SM-like Higgs boson in $ 123-127 {\rm GeV}$. Then we scrutinize the properties of
the surviving parameter space. Finally, we draw our conclusions in Sec. IV.

\section{The SM-like Higgs mass in the MSSM and NMSSM}

In the MSSM, the Higgs sector consists of two doublet fileds $H_u$ and $H_d$, which after
the electroweak symmetry breaking, result in five physical Higgs bosons: two CP-even scalars
$h$ and $H$, one CP-odd pseudoscalar $A$ and a pair of charged scalars $H^\pm$\cite{Haber}.
Traditionally, such a Higgs sector is described by the ratio of the Higgs vacuum expectation
values, $\tan \beta \equiv \frac{v_u}{v_d}$, and the mass of the pseudoscalar $m_A$. In most
of the MSSM parameter space, the lightest Higgs boson $h$ has largest coupling to vector
bosons (i.e. the so-called SM-like Higgs boson), and for moderate $\tan \beta$ and large
$m_A$ its mass is given by \cite{Carena:2011aa}
\begin{equation}\label{mh}
 m^2_{h}  \simeq M^2_Z\cos^2 2\beta +
  \frac{3m^4_t}{4\pi^2v^2} \ln\frac{M^2_S}{m^2_t} +
\frac{3m^4_t}{4\pi^2v^2}\frac{X^2_t}{M_S^2} \left( 1 - \frac{X^2_t}{12M^2_S}\right),
\end{equation}
where $v=174 {\rm ~GeV}$, $M_S = \sqrt{m_{\tilde{t}_1}m_{\tilde{t}_2}}$ with $m_{\tilde{t}_1}$
and $m_{\tilde{t}_2}$  being the stop masses, $X_t \equiv A_t - \mu \cot \beta$ with $A_t$
denoting the trilinear Higgs-stop coupling and $\mu$ being the Higgsino mass parameter.
Obviously, the larger $\tan \beta$ or $M_S$ is, the heavier $h$ becomes, and for given $M_S$,
$m_h$ reaches its maximum when $X_t/M_S=\sqrt{6}$, which corresponds to the so-called
$m_h^{max}$ scenario.

About Eq.(\ref{mh}), three points should be noted \cite{Carena:2011aa}.
The first is this equation is only valid for small splitting between
$m_{\tilde{t}_1}$ and $m_{\tilde{t}_2}$. In case of large splitting,
generally $X_t/M_S > \sqrt{6}$ is needed to maximize $m_h$.
The second is $m_h^2$ in Eq.(\ref{mh}) is symmetric with respect to the
sign of $X_t$. This behavior will be spoiled once higher order corrections
are considered, and usually a larger $m_h$ is achieved for positive $A_t M_3$
with $M_3$ being gluino soft breaking mass. And the last is in Eq.(\ref{mh}),
we do not include the contributions from the sbottom and slepton sectors.
Such contributions are negative and become significant only for large $\tan \beta$.

Compared with the MSSM, the Higgs sector in the NMSSM is rather complex, which can
be seen from its superpotential and the corresponding soft-breaking terms given by
\cite{NMSSM1}
\begin{eqnarray} \label{NMSSM}
 W_{\rm NMSSM}&=&W_F + \lambda\hat{H_u} \cdot \hat{H_d} \hat{S}
 + \frac{1}{3}\kappa \hat{S^3},\\
V_{\rm soft}^{\rm NMSSM}&=&\tilde m_u^2|H_u|^2 + \tilde m_d^2|H_d|^2
+ \tilde m_S^2|S|^2 +(A_\lambda \lambda SH_u\cdot H_d
+\frac{A_\kappa}{3}\kappa S^3 + h.c.).
\end{eqnarray}
Here $W_F$ is the superpotential of the MSSM without the $\mu$ term,
the dimensionless parameters $\lambda$ and $\kappa$ are the coefficients
 of the Higgs self couplings, and $\tilde{m}_{u}$, $\tilde{m}_{d}$, $\tilde{m}_{S}$,
$A_\lambda$ and $A_\kappa$ are the soft-breaking parameters.

After the electroweak symmetry breaking, the three soft breaking masses squared for
$H_u$, $H_d$ and $S$ can be expressed in terms of their VEVs (i.e. $v_u$, $v_d$ and $s$)
through the minimization conditions of the scalar potential.
So in contrast to the MSSM where there are only two parameters in the Higgs sector,
the Higgs sector of the NMSSM is described by six parameters \cite{NMSSM1}:
\begin{eqnarray}
\lambda, \quad \kappa, \quad M_A^2= \frac{2 \mu (A_\lambda + \kappa s)}{\sin 2 \beta},
\quad A_\kappa, \quad \tan \beta=\frac{v_u}{v_d}, \quad \mu = \lambda s.
\end{eqnarray}
The Higgs fields can be written in the following form:
\begin{eqnarray}
H_1 = \left ( \begin{array}{c} H^+ \\
       \frac{S_1 + i P_1}{\sqrt{2}}
        \end{array} \right),~~
H_2 & =& \left ( \begin{array}{c} G^+
            \\ v + \frac{ S_2 + i G^0}{\sqrt{2}}
            \end{array} \right),~~
H_3  = s +\frac{1}{\sqrt{2}} \left(  S_3 + i P_2 \right),
\end{eqnarray}
where $H_1 = \cos \beta H_u -  \varepsilon \sin \beta H_d^\ast$,
$H_2 = \sin \beta H_u + \varepsilon \cos \beta H_d^\ast$ with $\varepsilon_{12}=\varepsilon_{21}=-1$
and $\varepsilon_{11}=\varepsilon_{22}=0$,
$G^+$ and $G^0$ are Goldstone bosons and $v=\sqrt{v_u^2+v_d^2}$.
In the CP-conserving NMSSM,
the fields $S_1$, $S_2$ and $S_3$ mix to form three physical CP-even
Higgs bosons, and $P_1$ and  $P_2$ mix to form two physical CP-odd
Higgs bosons. Obviously, the field $H_2$ corresponds to the SM Higgs field,
and the scalar $h$ with largest $S_2$ component is called the SM-like Higgs boson.

Under the basis ($S_1$, $S_2$, $S_3$), the elements of the mass matrix for $S_i$ fields at
tree level are given by \cite{Kang:2012tn,Miller:2003ay}
\begin{eqnarray}
{\cal M}^2_{11}&=&M_A^2+(m_Z^2-\lambda^2v^2)\sin^22\beta,\\
{\cal M}^2_{12}&=&-\frac{1}{2}(m_Z^2-\lambda^2v^2)\sin4\beta,\\
{\cal M}^2_{13}&=&-(M_A^2\sin2\beta+\frac{2\kappa\mu^2}{\lambda})\frac{\lambda v}{\mu}\cos2\beta,\\
{\cal M}^2_{22}&=&m_Z^2\cos^22\beta+\lambda^2v^2\sin^22\beta,\label{22}\\
{\cal M}^2_{23}&=& 2 \lambda \mu v \left[1 - (\frac{M_A \sin 2\beta}{2 \mu} )^2
-\frac{\kappa}{2 \lambda}\sin2\beta\right],\\
{\cal M}^2_{33}&=& \frac{1}{4} \lambda^2 v^2 (\frac{M_A \sin 2\beta}{\mu})^2
+ \frac{\kappa\mu}{\lambda} (A_\kappa +  \frac{4\kappa\mu}{\lambda} )
 - \frac{1}{2} \lambda \kappa v^2 \sin 2 \beta, \label{33}
\end{eqnarray}
where ${\cal M}^2_{22}$ is nothing but $m_h^2$ at tree level without considering the mixing
among $S_i$, and its second term $\lambda^2v^2\sin^22\beta$ originates from the coupling
$\lambda\hat{H_u} \cdot \hat{H_d} \hat{S}$ in the superpotential.
For such a complex matrix, it is useful to consider two scenarios for understanding the
results:
\begin{itemize}
\item{Scenario I:} $\lambda,\kappa \to 0$ and $\mu$ is fixed.
In this limit, since ${\cal M}^2_{13}, {\cal M}^2_{23} \simeq 0$,
the singlet field $S_3$ is decoupled from the doublet fields, and the MSSM mass matrix
is recovered for the ($S_1$,$S_2$) system.  This scenario indicates that, even for moderate
$\lambda$ and $\kappa$, $m_h$ should change little from its MSSM prediction.
So in order to show the difference of the two
models in predicting $m_h$, we are more interested in large $\lambda$ case.
Especially we will mainly discuss $\lambda > M_Z/v \simeq 0.53$ case, where the
tree level contributions to $m_h^2$, i.e. ${\cal M}^2_{22}$, are maximized for
moderate values of $\tan \beta$ rather than by large values of $\tan \beta$ as
in the MSSM.
\item{Scenario II:} ${\cal M}^2_{11}\gg {\cal M}^2_{22}\gg {\cal M}^2_{12}$ and
$({\cal M}^2_{11} - {\cal M}^2_{33}) \gg {\cal M}^2_{13}$, which can be easily realized
for a large $M_A^2$. In this limit, $S_1$ is decoupled from
the ($S_2$,$S_3$) system and the properties of $m_h$ can be qualitatively understood by
the $2\times 2$ matrix \cite{Kang:2012tn}
\begin{eqnarray}  \label{Re-M}
 \tilde{M}^2 = \left ( \begin{array}{cc} {\cal M}^2_{22}+\delta^2 &{\cal M}^2_{23} \\
       {\cal M}^2_{23} &{\cal M}^2_{33}-\Delta^2
        \end{array} \right),
\end{eqnarray}
where $\delta^2$ denotes the radiative corrections to $m_h$ with its form given by
the last two terms of Eq.(\ref{mh}), and $\Delta^2$ represents the potentially important
effect of ($S_1$,$S_3$) mixing on $\tilde{M}^2_{22}$. This matrix indicates that
 the ($S_2$,$S_3$) mixing can push  $m_h$ up once $\tilde{M}^2_{11} > \tilde{M}^2_{22}$, and
such effect is maximized for $\tilde{M}_{11}^2 $ slightly
larger than  $\tilde{M}_{22}^2 $ and at the same time $\tilde{M}_{12}^4 = {\cal{M}}^4_{23}$
slightly below $\tilde{M}_{11}^2 \tilde{M}_{22}^2$
(larger $\tilde{M}_{12}$ will destabilize the vacuum) \cite{Kang:2012tn}.
Obviously, in this push-up case, $h$ is the next-to-lightest
CP-even Higgs boson and the larger the ($S_2, S_3$) mixing is, the heavier $h$ becomes.
Alternatively, the mixing can pull $m_h$ down on the condition of
$\tilde{M}^2_{11} < \tilde{M}^2_{22}$, which occurs for large $\kappa \mu$
(for $M_A \sin 2 \beta/\mu \sim 2$, see discussion below)
as indicated by the expression of ${\cal M}^2_{33}$ and the positiveness of
$\Delta^2$. Here we remind that, due to the extra contribution
$\lambda^2v^2\sin^22\beta$ to $m_h^2$ at tree level, $m_h$ in the pull-down
case may still be larger than its MSSM prediction for a certain $\delta^2$.

Since our results presented below is approximately described by scenario II, we
now estimate the features of its favored region to predict $m_h \simeq 125 {\rm GeV}$.
First, since $\tilde{M}^2_{11} \sim {\cal{O}}(100^2) {\rm GeV^2}$,
${\cal M}^2_{11} \gg {\cal M}^2_{22}$ implies that  $M_A \gtrsim {\cal{O}}(300) {\rm GeV}$.
Numerically, we find $M_A \gtrsim {\cal{O}}(300) {\rm GeV}$ for the push-up case, and
$M_A \gtrsim {\cal{O}}(500) {\rm GeV}$ for the pull-down case (see Fig.\ref{fig7}).
Second, $\tilde{M}_{12}^2={\cal{M}}^2_{23}$ must be relatively small, which implies that
$M_A \sin 2 \beta/\mu \sim 2$ for $\lambda > 0.53$.
This can be understood as follows. In the push-up case, since
$\tilde{M}^2_{22} < \tilde{M}^2_{11} \sim  {\cal{O}}(100^2) {\rm GeV^2}$,
the condition ${\cal{M}}^4_{23} < \tilde{M}^2_{11}  \tilde{M}^2_{22} $
(for vacuum stability) has limited the
size of ${\cal{M}}^2_{23}$. While in the pull-down case, a very large ${\cal{M}}^2_{23}$
will suppress greatly $m_h$ to make it difficult to reach $125 {\rm GeV}$, and this
in return limits the size of ${\cal{M}}^2_{23}$. Given that $\mu \gtrsim 100 {\rm ~GeV}$
as required by the LEP bound
on chargino mass and that a larger $\mu$ is favored for the pull-down scenario,
one can infer that the value of $M_A \sin 2 \beta/\mu $ should be around 2 after
considering that the third term in ${\cal{M}}^2_{23}$ is less important.
Numerically speaking, we find $|{\cal{M}}^2_{23}/(2 \lambda \mu v)| \leq 0.2 $
and $1.4 \leq M_A \sin 2 \beta/\mu \leq 2$ (see Fig.\ref{fig7}).
Lastly,  light stops may be possible in the NMSSM with large $\lambda$ to
predict $m_h \simeq 125 {\rm GeV}$.
To see this, we consider the parameters $\lambda =0.7 $ and $\tan \beta=1.5$, and
we get $\delta^2/125^2 \sim 5\%$ without considering the mixing effect to
predict $m_h \simeq 125 {\rm GeV}$. This is in sharp contrast
with $\delta^2/125^2 \sim 55\%$ in MSSM for $\tan \beta=5$.
\end{itemize}

In this work we use the package NMSSMTools \cite{NMSSMTools} to calculate the Higgs masses
and mixings, which includes the dominant one-loop and leading logarithmic two-loop corrections
to ${\cal{M}}^2$. We checked our MSSM results of $m_h$ by using the code FeynHiggs\cite{FeynHiggs}
and found the results given by  NMSSMTools and  FeynHiggs are in good agreement (
for $m_h \sim 125 GeV$ they agree within $0.5 GeV$ for same MSSM parameters).

\section{Numerical Results and discussions}

In this work, we scan the parameters of the models and investigate the samples that predict
$123 {\rm GeV} \leq m_h \leq 127 {\rm GeV}$ and at the same time survive
the following constraints \cite{NMSSMTools}: (1) The constraint from the
LHC search channel $p p \to H \to 2 \tau$ for non-standard Higgs boson.
(2) The limits from the LEP and the Tevatron on the masses of sparticles
as well as on the neutralino pair productions. (3) The constraints from
B-physics, such as $B\to X_s \gamma$, the latest experimental result of $B_s\to\mu^+\mu^-$,
$B_d \to X_s \mu^+ \mu^-$, $B^+ \to \tau^+ \nu$ and the mass differences $\Delta M_d$ and $\Delta M_s$.
(4) The indirect constraints from the electroweak precision observables
such as $\sin^2 \theta_{eff}^{\ell}$, $\rho_{\ell}$ and $M_W$,
and their combinations $\epsilon_i (i=1,2,3)$ \cite{Altarelli}.
We require $\epsilon_i$ to be compatible with the LEP/SLD data at $95\%$ confidence level.
We also require the SUSY prediction of the observable
$R_b$ ($\Gamma (Z \to \bar{b} b) / \Gamma ( Z \to {\rm hadrons} )$) to be within
the $2 \sigma$ range of its experimental value \cite{cao-zbb}.
(5) The constraints from the muon anomalous magnetic moment:
$a_\mu^{exp}-a_\mu^{SM} = (25.5\pm 8.2)\times 10^{-10}$ \cite{g-2}. We require
 the SUSY effects to explain the discrepancy at $2\sigma$ level.
(6) The dark matter constraints from WMAP relic density
(0.1053 $< \Omega h^2 <$ 0.1193) \cite{WMAP} and the direct search result
from XENON100 experiment (at $90\%$ C.L.) \cite{XENON}.
(7) For the NMSSM, we also require
the absence of a landau singularity below the GUT scale,
which implies $\lambda \lesssim 0.7$
for small $\kappa$ and $\kappa \lesssim 0.5 $ for $ \lambda > 0.53$ at weak scale.
In our calculation, we fix $m_t=172.9 {\rm ~GeV}$ and $f_{Ts}=0.02$
\cite{lattice} ($f_{Ts}$ denotes the strange quark fraction in the
proton mass), and use the package NMSSMTools to implement most of the constraints
and to calculate the observables we are interested in.

In our scan, we note that the soft parameters in the slepton sector can only affect
significantly the muon anomalous magnetic moment $a_\mu$, which will in return limit the
important parameter $\tan \beta$, so we assume them a common value $m_{\tilde{l}}$
and treat it as a free parameter. For the soft parameters in the first two generation squark sector,
due to their little effects on the properties of the Higgs bosons,
we fix them to be $1~{\rm TeV}$.  As for the gaugino masses, we assume the grand unification relation,
$3 M_1/5\alpha_1=M_2/\alpha_2=M_3/\alpha_3$ with $\alpha_i$ being the
fine structure constants of the different gauge groups, and treat $M_1$ as a free parameter. In order
to reduce free parameters, we also assume the unimportant parameters $M_{D_3}$ and $A_b$ to satisfy
$M_{D_3} = M_{U_3}$ and $A_b = A_t$.

\subsection{Implication of $m_h \simeq 125 {\rm GeV} $ in generic SUSY}
In order to study the implication of $m_h \simeq 125 {\rm GeV} $ in generic SUSY,
we relax the soft mass parameters to 5 TeV and
perform an extensive random scan over the following parameter regions:
\begin{eqnarray} \label{MSSM-scan}
&&1\leq\tan\beta \leq60,
~~90{\rm ~GeV}\leq m_A\leq 1 {\rm ~TeV},
~~100{\rm ~GeV}\leq \mu\leq 2 {\rm ~TeV}, \nonumber\\
&&100{\rm ~GeV}\leq  M_{Q_3},M_{U_3}\leq 5 {\rm ~TeV},
~~ |A_{t}|\leq 5 {\rm ~TeV},\nonumber\\
&&100{\rm ~GeV}\leq m_{\tilde{l}}\leq 1 {\rm ~TeV},
~~~50{\rm ~GeV}\leq M_1\leq 500 {\rm ~GeV},
\end{eqnarray}
for the MSSM, and
\begin{eqnarray}\label{NMSSM-scan}
&& 0<\lambda\leq 0.2,~ 0<\kappa \leq 0.7, ~90{\rm ~GeV}\leq M_A\leq 1 {\rm ~TeV},
~|A_{\kappa}|\leq 1{\rm ~TeV},
\nonumber\\
&& 100{\rm ~GeV}\leq M_{Q_3},M_{U_3}
\leq 5 {\rm ~TeV} ,~~|A_{t}|\leq 5 {\rm ~TeV}, \nonumber\\
&& 1\leq\tan\beta \leq 60,  ~~100{\rm
~GeV}\leq \mu,m_{\tilde{l}}\leq 1 {\rm ~TeV},~~50{\rm~GeV}\leq M_1\leq 500 {\rm ~GeV},
\end{eqnarray}
for the NMSSM. In our scan, we only keep the samples satisfying the requirements listed
in the text (including $ 123{\rm GeV} \leq m_h \leq 127 {\rm GeV} $). To show the differences
between the MSSM and the NMSSM, we also perform a scan similar to Eq.(\ref{NMSSM-scan})
except that we require $\lambda> m_Z/v \simeq 0.53$.
\begin{figure}[t]
\centering
\includegraphics[width=5.6cm]{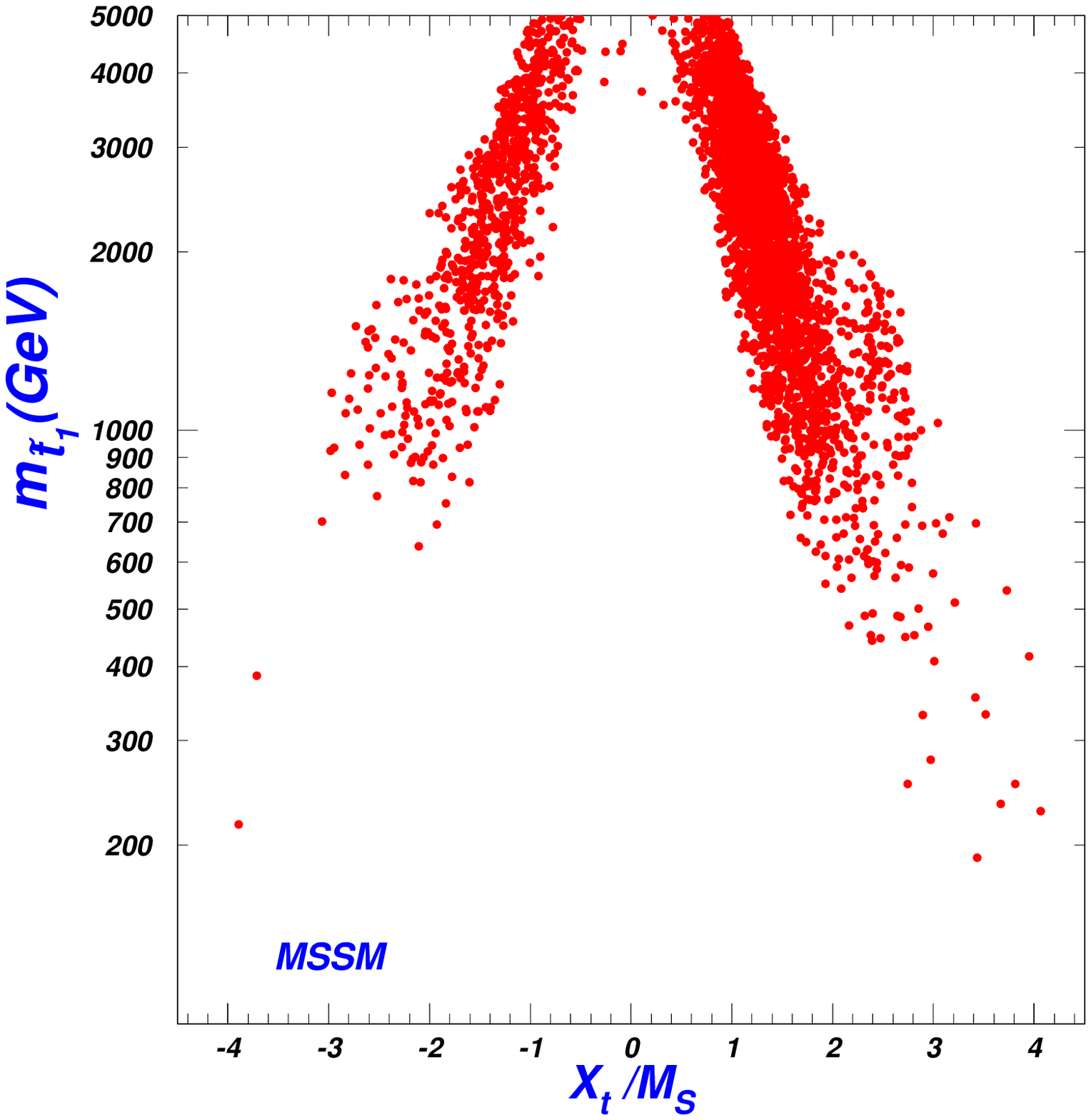}\hspace{-1cm}
\includegraphics[width=5.6cm]{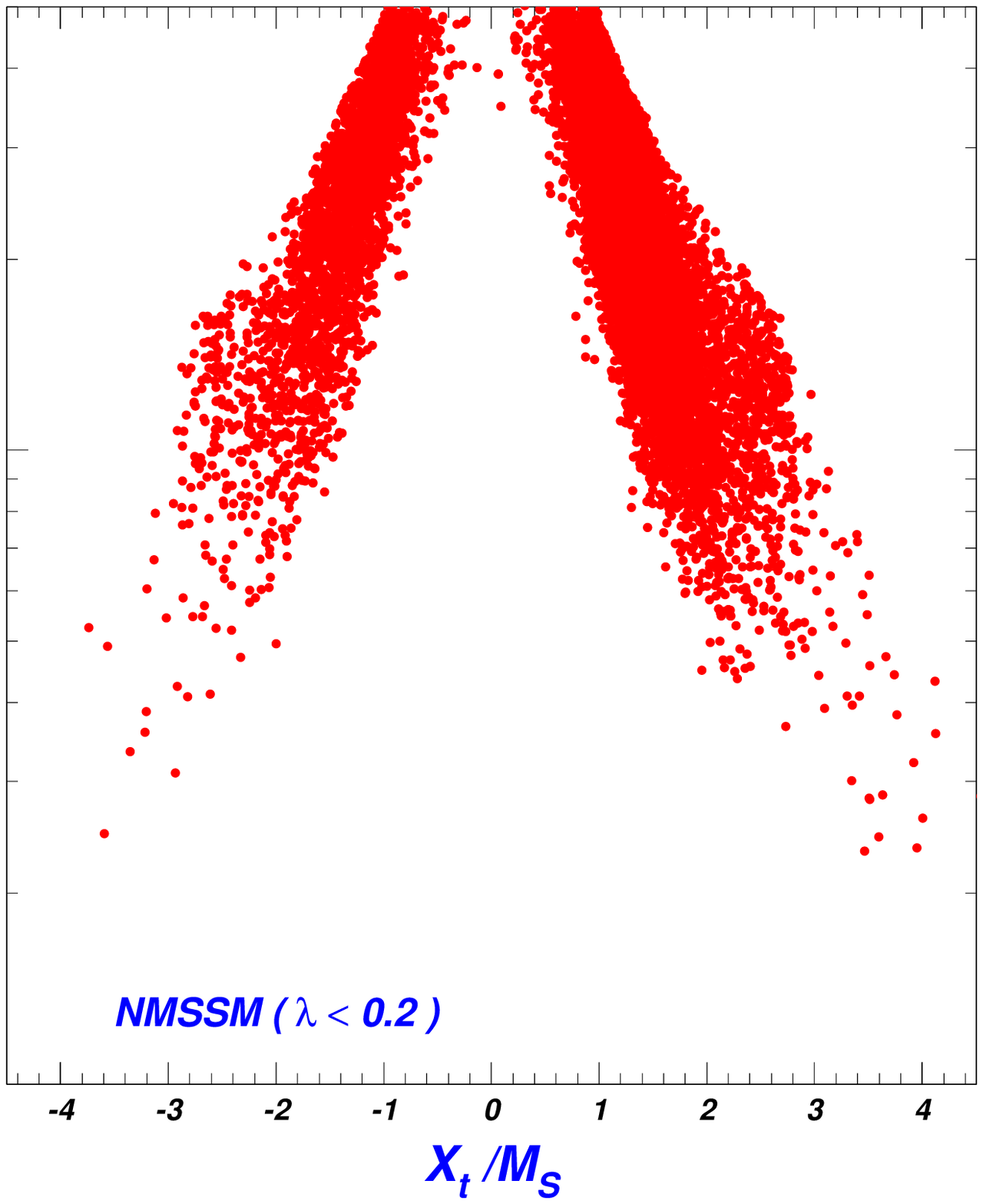}\hspace{-1cm}
\includegraphics[width=5.6cm]{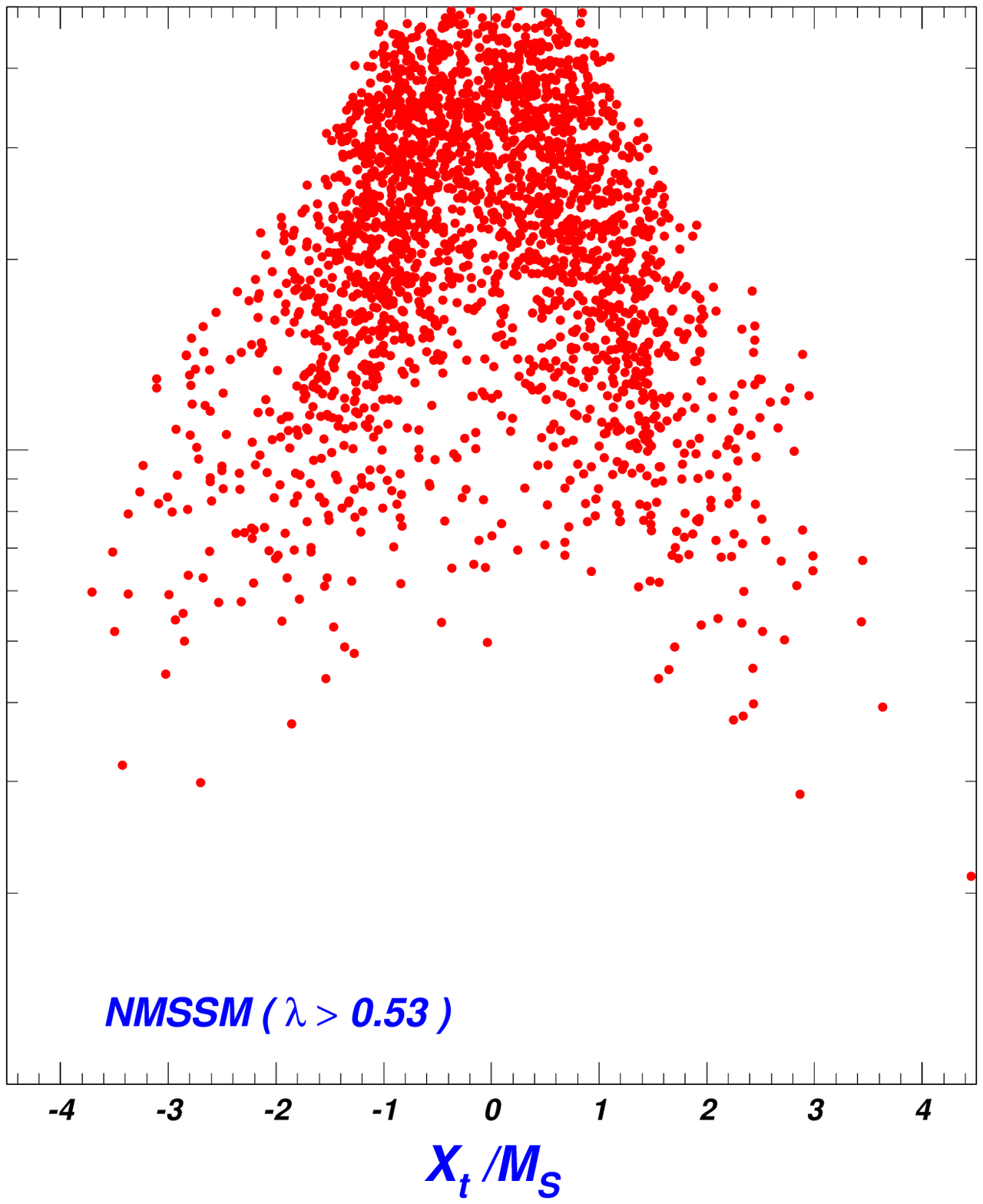}
\vspace*{-0.5cm}
\caption{The scatter plots of the samples in
the MSSM and NMSSM satisfying all the requirements (1-7)
listed in the text (including $ 123{\rm GeV} \leq m_h \leq 127 {\rm GeV} $),
projected in the plane of $m_{\tilde{t}_1}$ versus $X_t/M_{S}$ with
$M_S \equiv \sqrt{m_{\tilde t_1} m_{\tilde t_2}}$ and $X_t \equiv A_t - \mu \cot \beta$.}
\label{fig1}
\end{figure}
In Fig.\ref{fig1}, we show the correlation of the lighter top-squark mass
($m_{\tilde t_1}$) with the ratio $X_t/M_S$ ($M_S = \sqrt{m_{\tilde t_1} m_{\tilde t_2}}$)
for the surviving samples in the MSSM and NMSSM. As expected from Eq.(\ref{mh}), in order to predict
$m_h \simeq 125 {\rm GeV}$ in the MSSM, a large $X_t$ is needed for a moderate light $\tilde t_1$,
and with $\tilde{t}_1$ becoming heavy, the ratio $X_t/M_S$ decreases, but is unlikely
to vanish for $m_{\tilde t_1} < 5  {\rm TeV}$. These features are maintained for NMSSM
with $\lambda \leq 0.2 $ (see the middle panel) but changed for NMSSM with a large $\lambda$,
where $X_t$ can possibly vanish even for $m_{\tilde t_1} \sim 1 {\rm TeV}$.
Fig.\ref{fig1} also shows that a $\tilde{t}_1$ as light as $200{\rm GeV}$
is still able to give the required $m_h$. But in this case $X_t$ is large
($X_t/\sqrt{m_{\tilde{t}_1} m_{\tilde{t}_2}} > \sqrt{6}$), which leads to a large
mass splitting between two stops ($m_{\tilde{t}_2} \gg m_{\tilde{t}_1}$).
Note that a $\tilde{t}_1$ as light as $200 {\rm GeV}$ does not contradict the recent
SUSY search result of the LHC \cite{stop}.

Since in heavy SUSY the radiative correction $\delta^2$ is usually very large,
$m_h \simeq 125 {\rm GeV}$ is unlikely to impose tight constraints on other
parameters of the models. Considering heavy SUSY is disfavored by naturalness,
we in the following concentrate on the implication of $m_h \simeq 125 {\rm GeV}$ in
sub-TeV SUSY.

\subsection{Implication of $m_h \simeq 125 {\rm GeV} $ in sub-TeV SUSY}
In this section, we study the implication of  $m_h \simeq 125 {\rm GeV} $
in low energy MSSM and NMSSM. In order to illustrate the new features of the NMSSM,
we only consider the case with $\lambda > 0.53$. Our scans over the parameter
spaces are quite similar to those in Eq.(\ref{MSSM-scan}) and
Eq.(\ref{NMSSM-scan}) except that we narrow the ranges
of $M_{Q_3}$, $M_{U_3}$ and $A_t$ as follows:
\begin{eqnarray}
100{\rm ~GeV}\leq (M_{Q_3},M_{U_3})\leq 1 {\rm ~TeV} ,~~|A_{t}|\leq 3 {\rm ~TeV}.
\label{narrow}
\end{eqnarray}
\begin{figure}[htb]
\includegraphics[width=6.5cm]{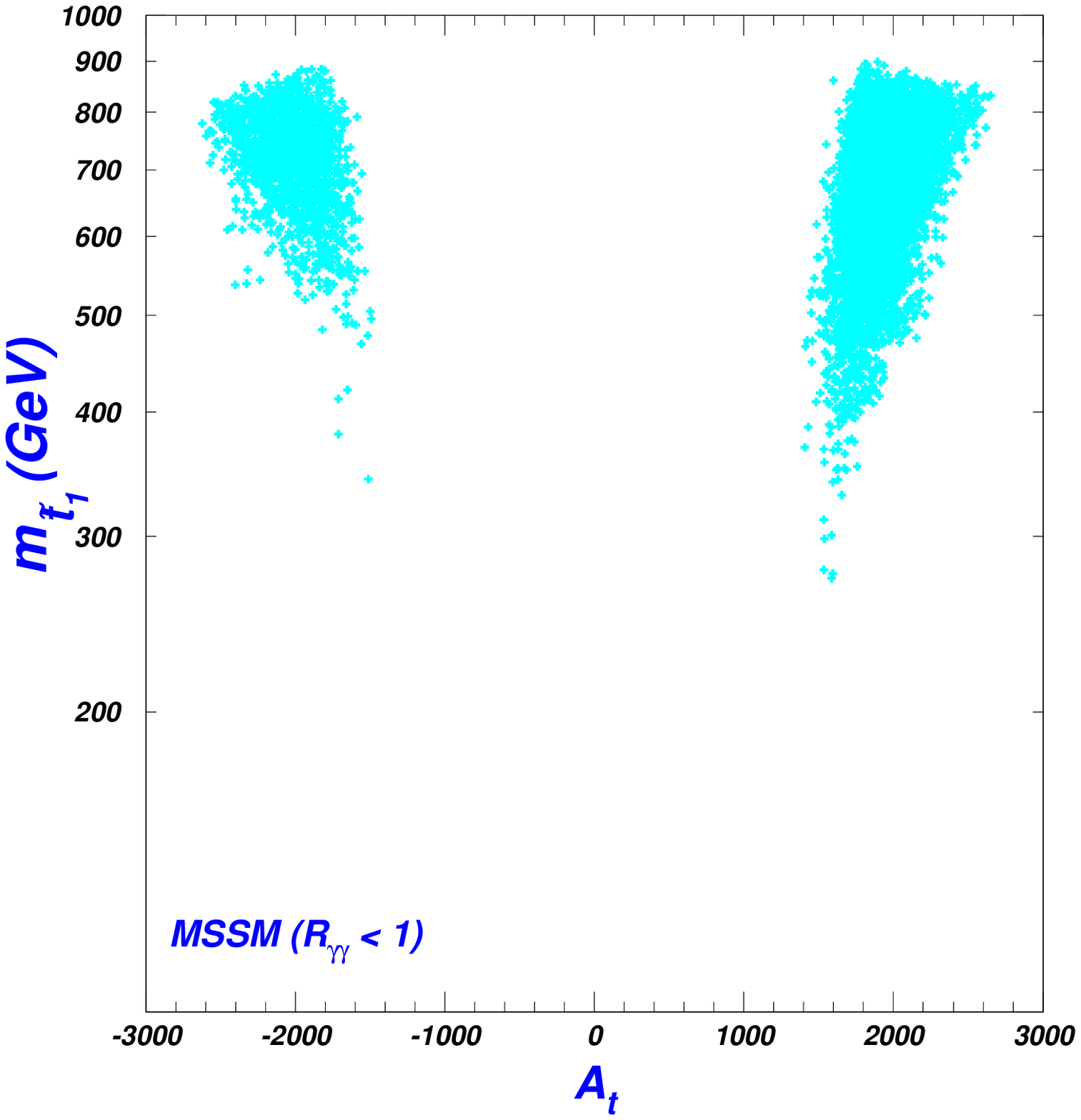}\hspace{0.2cm}
\includegraphics[width=6.5cm]{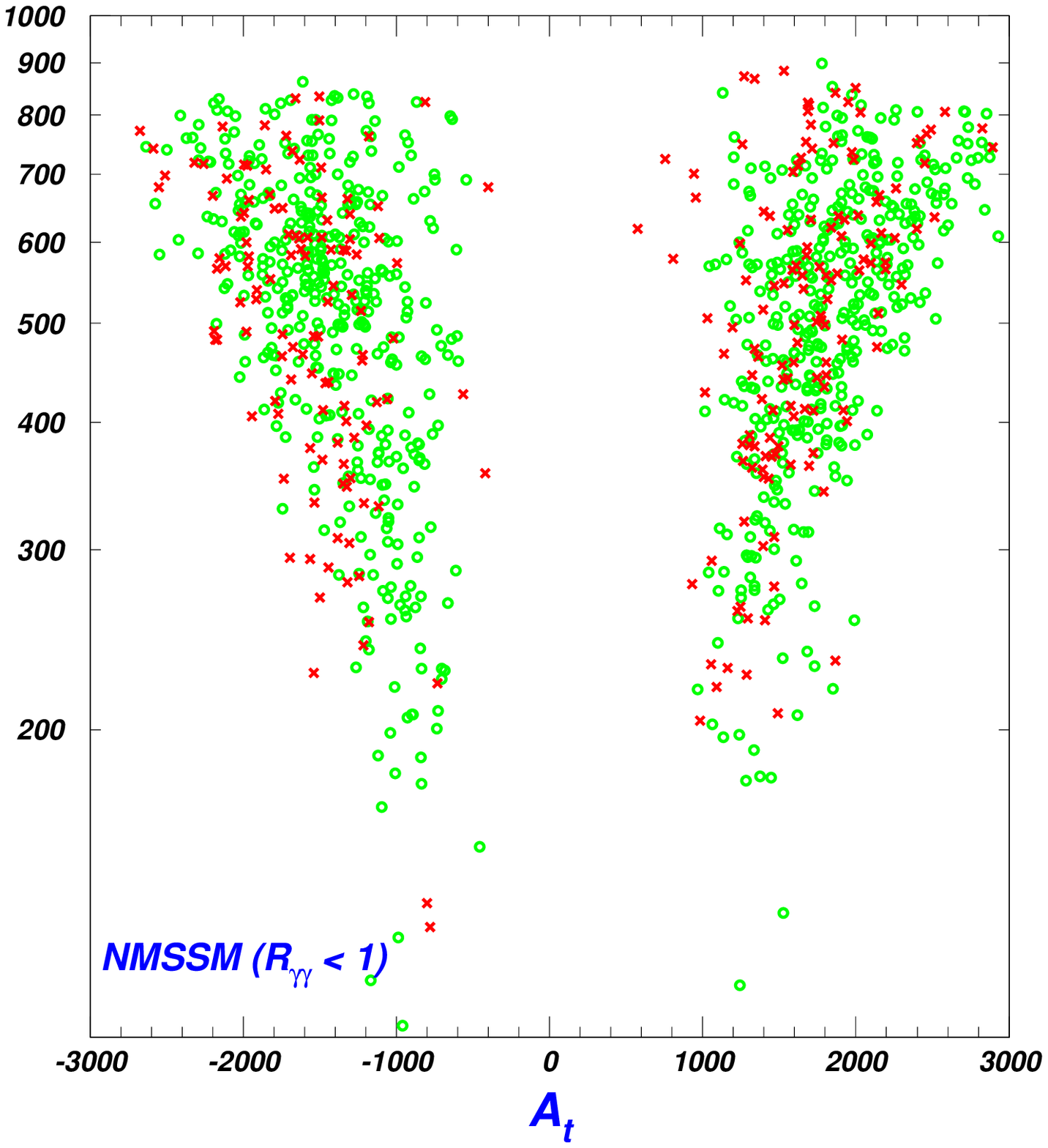}
    \includegraphics[width=6.5cm]{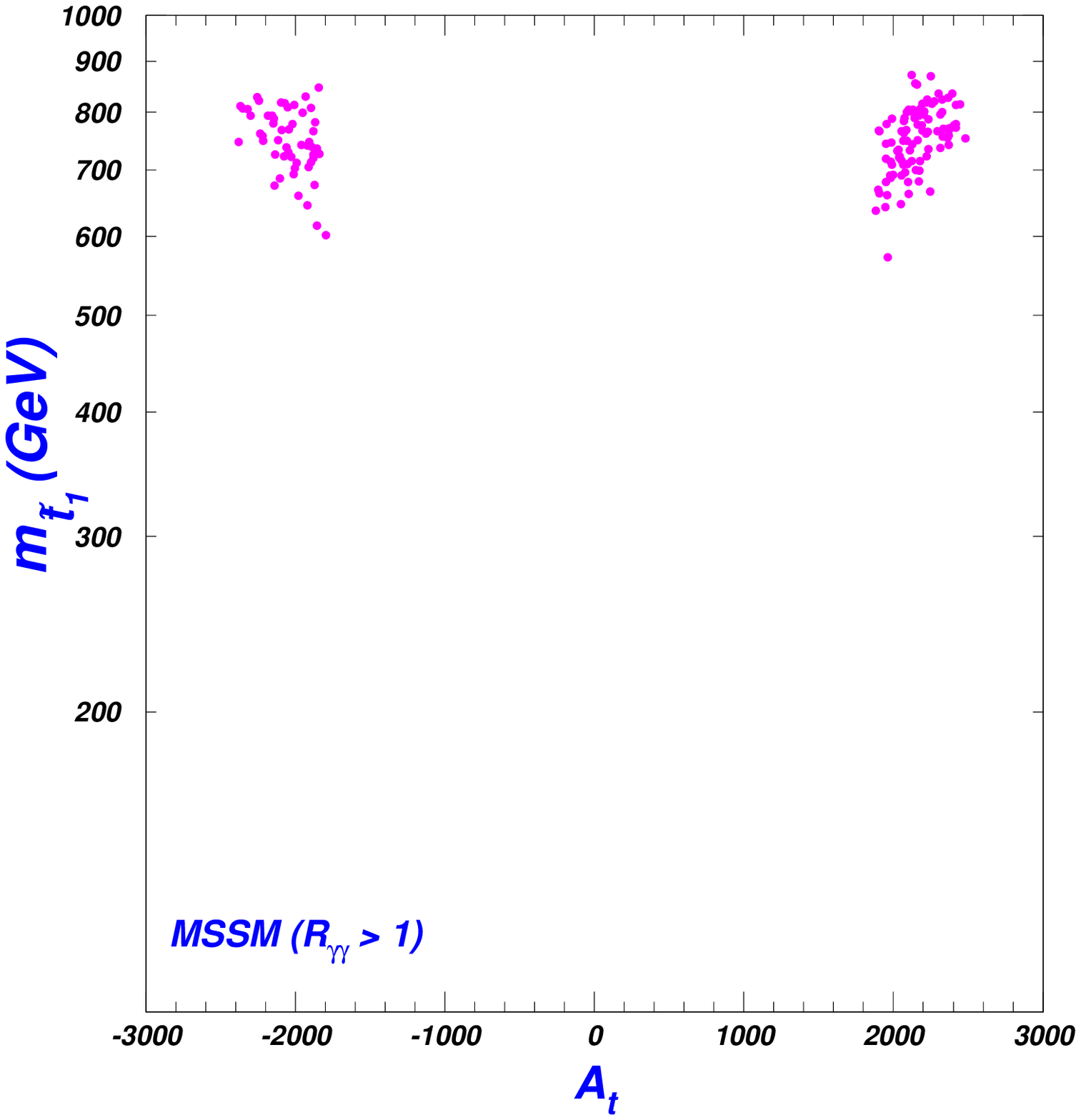}\hspace{0.2cm}
    \includegraphics[width=6.5cm]{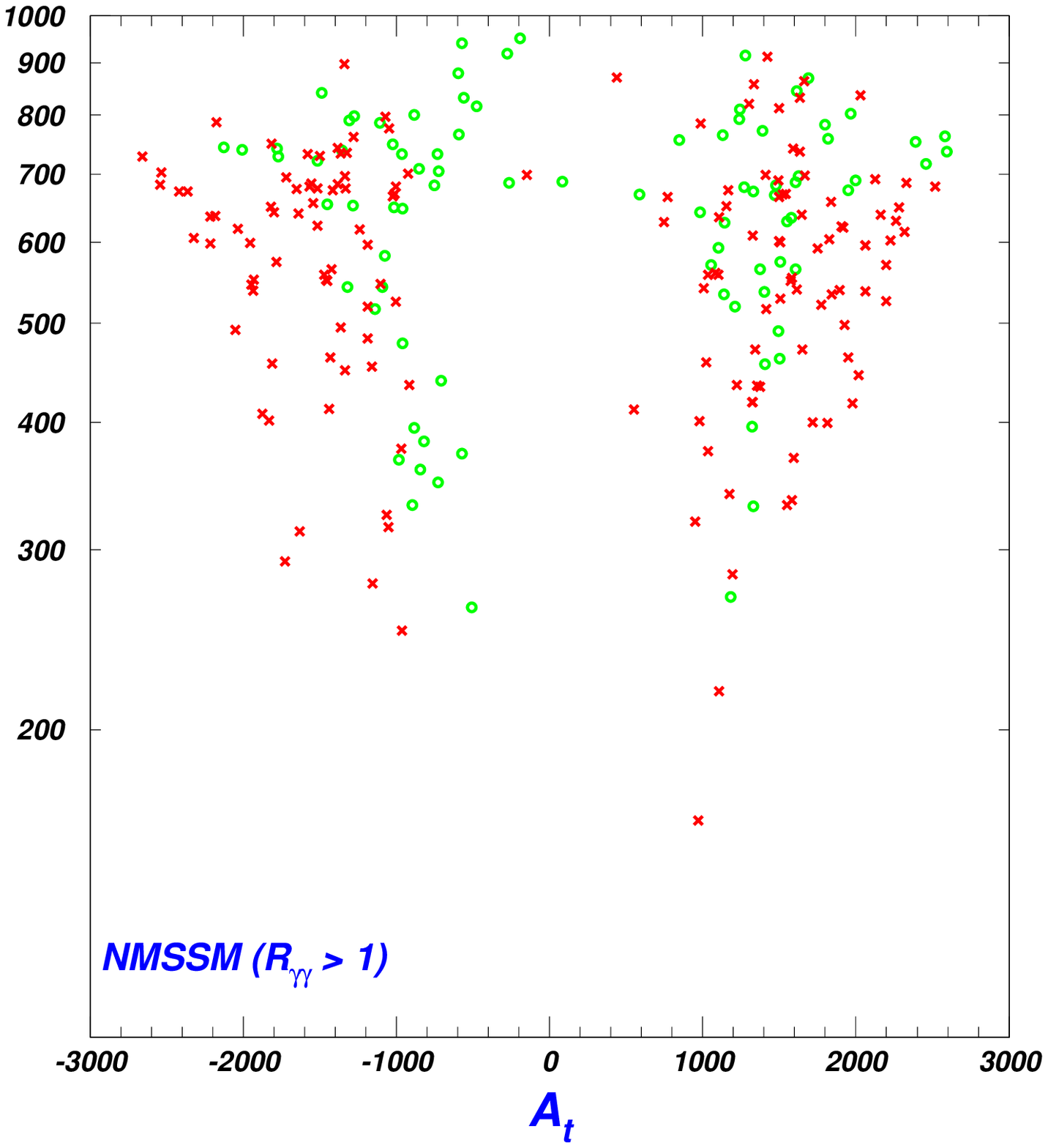}
\vspace*{-0.5cm}
\caption{Same as Fig.1, except that the samples are based on the narrowed
scan ranges of the soft masses shown in Eq.(\ref{narrow}) for both models
and the requirement $\lambda > 0.53$ for the NMSSM. Here the samples are
projected in the plane of $m_{\tilde{t}_1}$ versus $A_t$. The upper (lower) panels
correspond to $R_{\gamma \gamma} < 1$ ($R_{\gamma \gamma} > 1$)
with $R_{\gamma\gamma}\equiv \sigma_{SUSY} ( p p \to h \to
\gamma \gamma)/\sigma_{SM} ( p p \to h \to \gamma \gamma )$.
For the NMSSM results, the circles (green) denote the
case of the lightest Higgs boson being the SM-like Higgs (the so-called pull-down case),
and the times (red) denotes the case of
the next-to-lightest Higgs boson being the SM-like Higgs (the so-called push-up case). }
\label{fig2}
\end{figure}
In Fig.\ref{fig2} we project the surviving samples of the models in
the plane of $m_{\tilde{t}_1}$ versus $A_t$, showing the results
with $R_{\gamma \gamma} < 1$ and $R_{\gamma \gamma} >1 $ separately.
As we analyzed in Sec. II, the SM-like Higgs in the NMSSM may be either
the lightest Higgs boson (corresponding to the pull-down case) or the
next-to-lightest Higgs boson (the push-up case).
In the figure we distinguished these two cases.
We note that among the surviving samples
the number of the pull-down case is about twice
the push-up case.

Fig.\ref{fig2} shows that, in order to get $m_h \simeq 125 {\rm GeV}$ in the MSSM,
$m_{\tilde{t}_1}$ and $|A_t|$ must be larger than about $300 {\rm GeV}$ and
$1.5 TeV$ respectively, and the bounds are pushed up to $600 {\rm GeV}$ and
$1.8 TeV$ respectively for $R_{\gamma \gamma} > 1$. While in the NMSSM,
a $\tilde{t}_1$ as light as about $100 {\rm GeV}$ (in either the pull-down
or push-up case) is still able to predict  $m_h \simeq 125 {\rm GeV}$,
and even if one require $R_{\gamma \gamma} > 1$, a $\tilde{t}_1$
as light as about $200 {\rm GeV}$ is allowed.
The fact that the NMSSM allows a lighter $\tilde{t}_1$ than the MSSM
indicates that the NMSSM is more natural than the MSSM in light of
the LHC results.

\begin{table}[htb]
\caption{The ranges of the rescaled couplings $C_{hgg}\equiv C^{\rm SUSY}_{hgg}/ C^{\rm SM}_{hgg}$
and $C_{h\gamma \gamma}\equiv C^{\rm SUSY}_{h\gamma \gamma}/C^{\rm SM}_{h\gamma \gamma}$
predicted by the surviving samples of the two models.
The region of $C_{hgg}$($R_{\gamma\gamma} > 1$) is obtained by only considering the samples
with $R_{\gamma \gamma} >1$.}
 \begin{tabular}{|c|c|c|c|c|} \hline
 & $C_{hgg}$ & $C_{hgg}$($R_{\gamma\gamma} > 1$) & $C_{h\gamma\gamma}$ & $C_{h\gamma\gamma}$($R_{\gamma\gamma} > 1$)  \\
 \hline
MSSM  & 0.85 $\sim$ 0.99  & 0.95 $\sim$ 0.99 & 1 $\sim$ 1.25  & 1.05 $\sim$ 1.25
 \\  \hline
NMSSM & 0.3 $\sim$ 1 & 0.7 $\sim$ 1  & 0.7 $\sim$ 1.2  & 0.85 $\sim$ 1.05 \\ \hline
 \end{tabular}
 \end{table}

Since a light $\tilde{t}_1$ may significantly change the effective couplings $C_{hgg}$
and $C_{h\gamma \gamma}$, we present in Table I their predicted ranges
for the surviving samples. This table shows that $C_{hgg}$ is always reduced,
and for $m_{\tilde{t}_1} \sim 100 {\rm GeV}$ in the NMSSM, the reduction factor may
reach $70\%$. While $C_{h\gamma \gamma}$ exhibits quite strange behaviors: it is
enhanced in the MSSM, but may be either enhanced or suppressed in the NMSSM.
This is because,
unlike $C_{hgg}$ which is affected only by squark loops, $C_{h\gamma \gamma}$
gets new physics contributions from loops mediated by charged Higgs boson, charginos,
sleptons and also squarks, and there exists cancelation among
different loops. As will be shown below, the current experiments can not
rule out light sparticles like $\tilde{\tau}_1$ and chargino. Although
the contributions of these particles to $C_{h\gamma \gamma}$ are
far smaller than the $W$ loop contribution,
they may still alter the coupling significantly.

\begin{figure}[t]
\centering
\includegraphics[width=7cm]{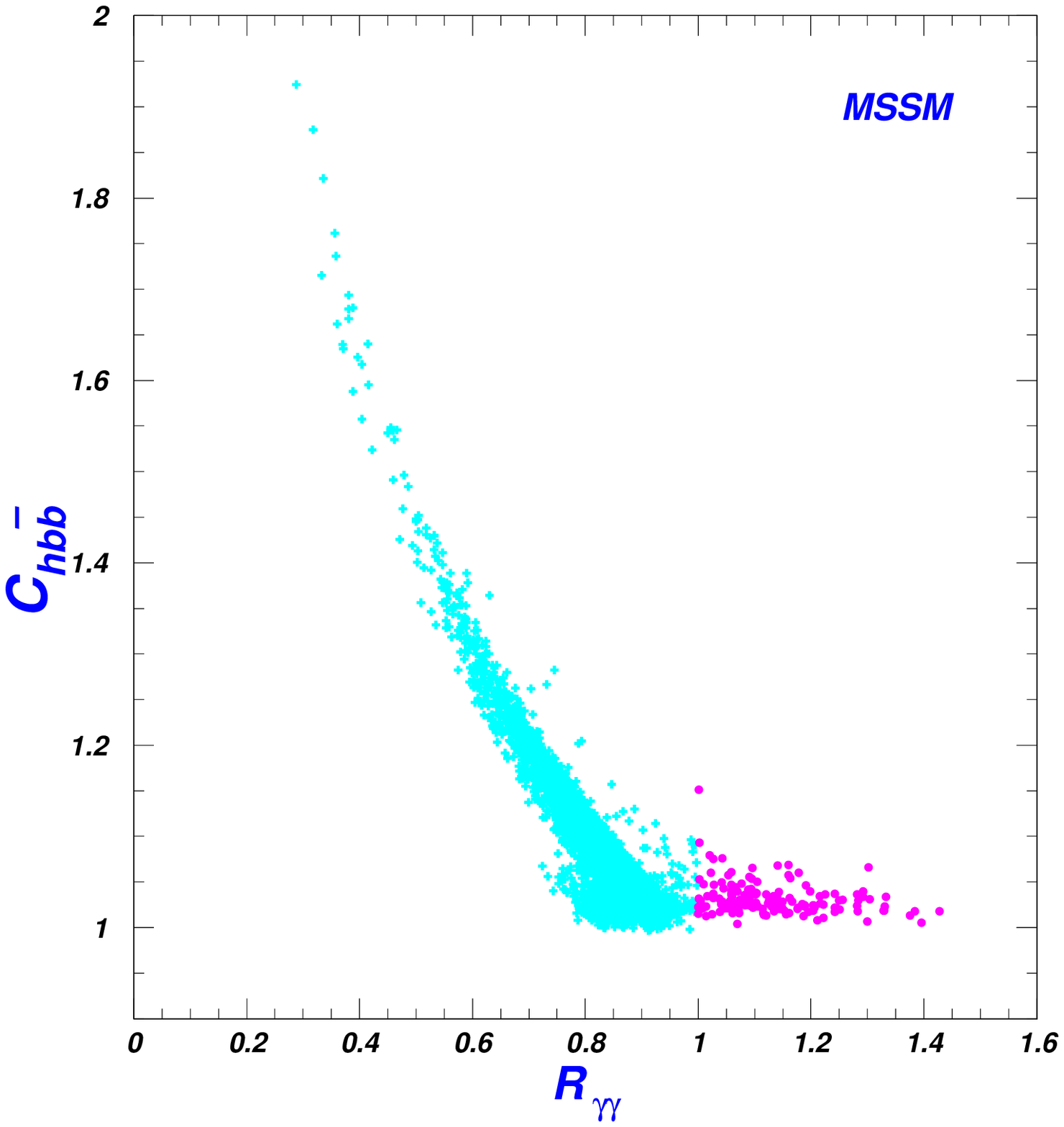}\hspace{0.2cm}
\includegraphics[width=7cm]{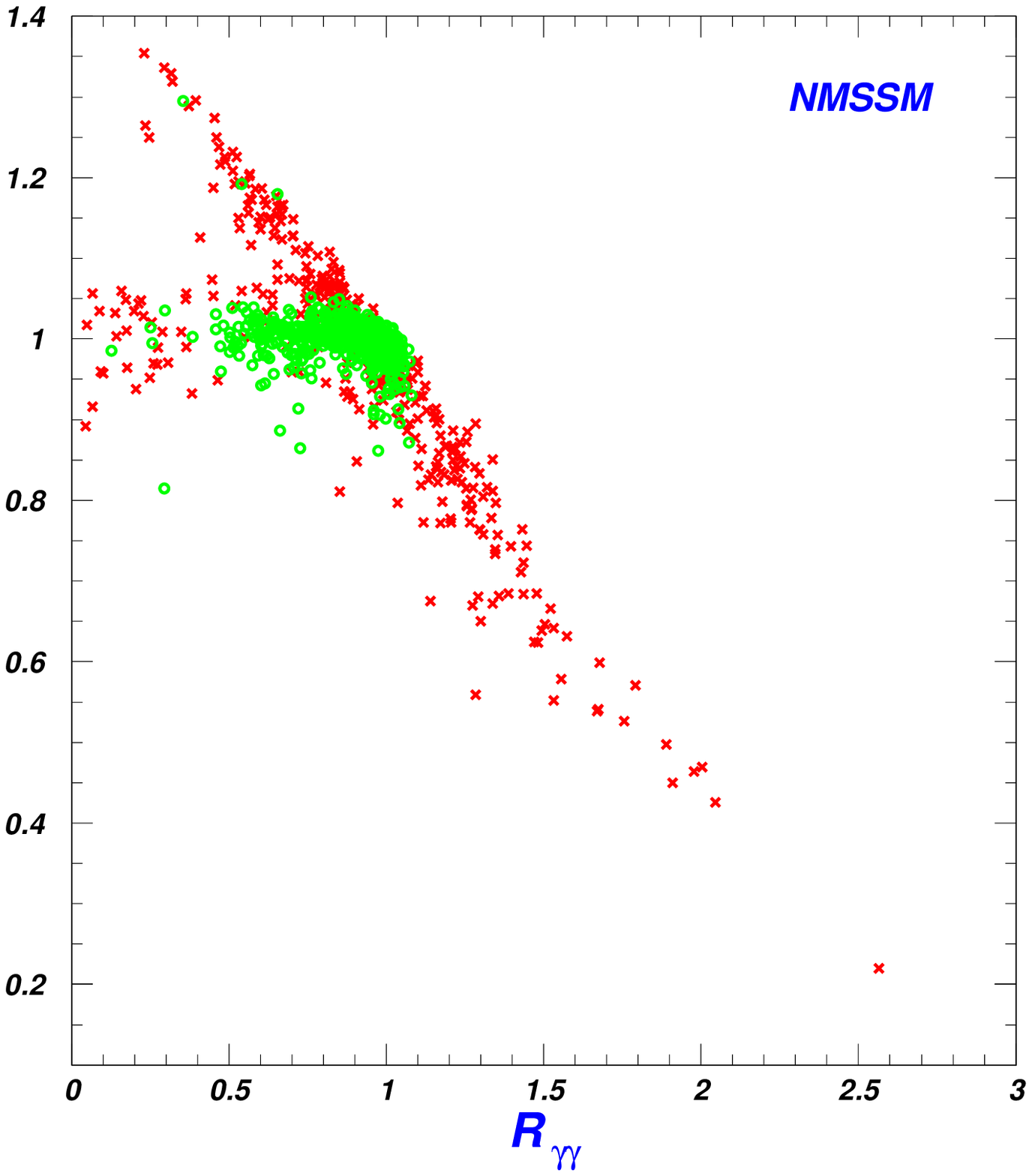}
\vspace*{-0.5cm}
\caption{Same as Fig.\ref{fig2}, but showing the dependence of the
di-photon signal rate $R_{\gamma\gamma}$
on the effective $h b\bar{b}$ coupling
$C_{h b\bar{b}}\equiv C^{\rm SUSY}_{h b\bar{b}}/C^{\rm SM}_{h b\bar{b}}$.}
\label{fig3}
\end{figure}
Since the di-photon signal is the most important discovery channel for the Higgs boson
around $125 {\rm GeV}$, it is useful to study its rate carefully.
From Eq.(\ref{definition1}) one can learn that the rate
is affected by $C_{hgg}$ and $C_{h\gamma \gamma}$ discussed above, and also by the total width of
$h$ (or more basically by the $hb\bar{b}$ coupling since $b \bar{b}$ is the dominant decay
of $h$). The importance of $hb\bar{b}$ coupling on the di-photon rate was recently emphasized
in \cite{diphoton3}. Here we restrict our study to the $m_h \simeq 125 {\rm GeV}$ case.
In Fig.\ref{fig3} we show the dependence of the di-photon signal rate $R_{\gamma\gamma}$
on the effective $h b\bar{b}$ coupling (including the potentially large
SUSY corrections \cite{Carena-eff,diphoton3}) normalized by its SM value.
This figure indicates that although the rate is suppressed for most of
the surviving samples in both models, there still exist some samples
with enhanced rate, especially the NMSSM is more likely to push up
the rate than the MSSM.
This feature can be understood as follows.
In SUSY, $R_{\gamma \gamma} > 1$ requires approximately
the combination $C_{hgg} C_{h\gamma \gamma}/C_{hb\bar{b}}$ to exceed 1.
For the MSSM, given $C_{hgg} < 1$ and $C_{h b \bar{b}} \geq 1$ for nearly all
the cases, this condition is not easy to satisfy.
While in the NMSSM, due to the mixing between the doublet field $S_2$
and the singlet field $S_3$, $C_{h b \bar{b}} < 1$ is possible once the singlet
component in $h$ is significant, which is helpful to enhance the combination.
In fact, we analyzed carefully the $R_{\gamma \gamma } > 1$ cases and found
they are characterized by $C_{hgg}, C_{hb\bar{b}} \simeq 1 $ and
$C_{h\gamma \gamma} > 1$ in the MSSM, and by $C_{hb\bar{b}} < 1$ in the NMSSM.
In other words, it is the enhanced $h\gamma \gamma$ coupling
(reduced total width of $h$) that mainly push up the di-photon
rate in the MSSM (NMSSM) to exceed its SM prediction.
Fig.\ref{fig3} also indicates that the pull-down case in the NMSSM is
less effective in reducing $C_{hb\bar{b}}$ and thus can hardly
enhance the di-photon rate. This is because
in the push-up case, both $\tilde{M}^2_{11}$ and $\tilde{M}^2_{22}$ in
Eq.(\ref{Re-M}) are moderate and often comparable,
which are helpful to enhance the ($S_2,S_3$) mixing.
Finally, we note that in some rare cases of the NMSSM
the ratio $R_{\gamma \gamma}$ may be very small even for $C_{hb\bar{b}} < 1$.
This is because there exists very light Higgs boson so that $h$ decays
dominantly into them.

\begin{figure}[t]
\centering
\includegraphics[width=7cm]{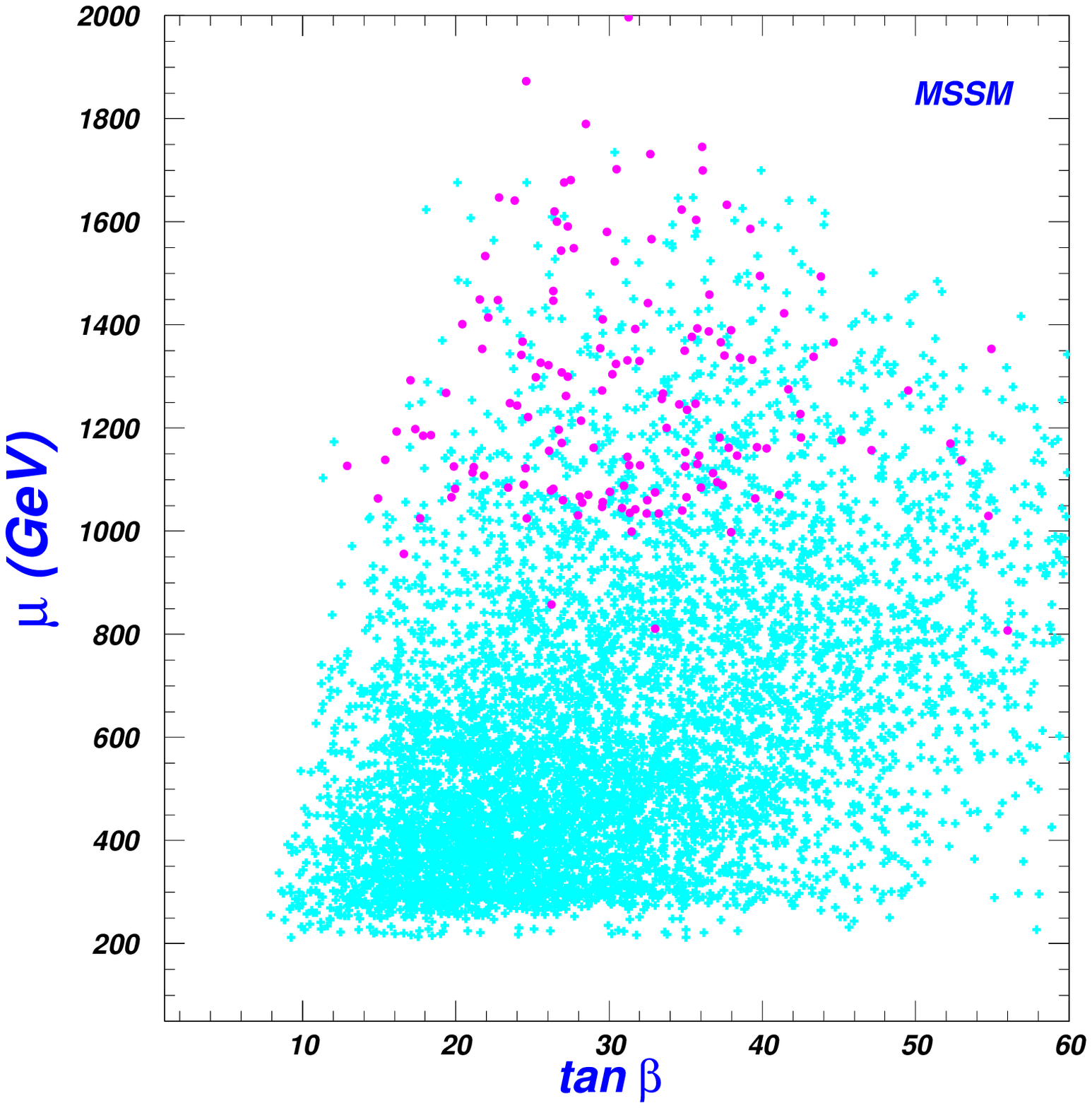}\hspace{0.5cm}
\includegraphics[width=7cm]{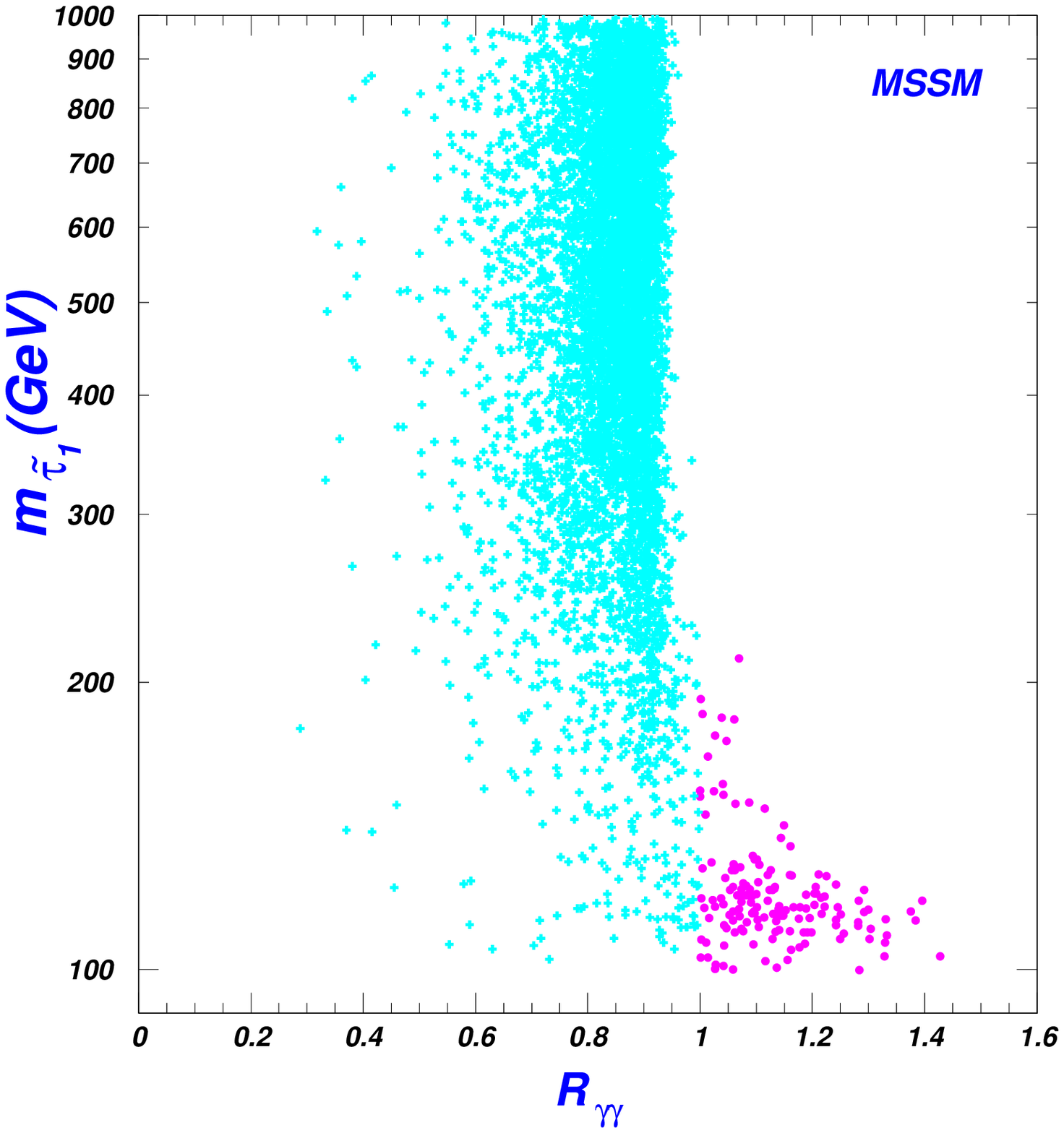}
\vspace*{-0.5cm}
\caption{Same as Fig.\ref{fig2}, but only for
the MSSM, projected in the planes of $\tan\beta$ versus $\mu$ and
$R_{\gamma \gamma}$ versus $m_{\tilde{\tau}_1}$.}
\label{fig4}
\end{figure}
In order to further clarify the reason for the enhancement of the di-photon rate in the MSSM,
we scrutinize carefully the parameters of the model and find
that the samples with $R_{\gamma \gamma} > 1$ correspond to the case
with a large $\mu \tan \beta$ and $m_{\tilde{\tau}_1}<200 {\rm GeV}$,
which is illustrated in Fig.\ref{fig4}. This means that the stau
loop plays an important role in enhancing $C_{\gamma \gamma}$. We note
that similar conclusion was recently achieved in \cite{Carena:2011aa}, but
in that work the authors did not consider the tight experimental constraints.
From Fig.\ref{fig4} we also note that $R_{\gamma \gamma} < 0.95$ is predicted
in the MSSM with $m_{\tilde{\tau}_1} > 250 {\rm GeV}$. So future precise
measurement of $R_{\gamma \gamma}$ and $ m_{\tilde{\tau}_1}$ may be utilized
to verify the correctness of the MSSM.

\begin{figure}[t]
\centering
\includegraphics[width=7cm]{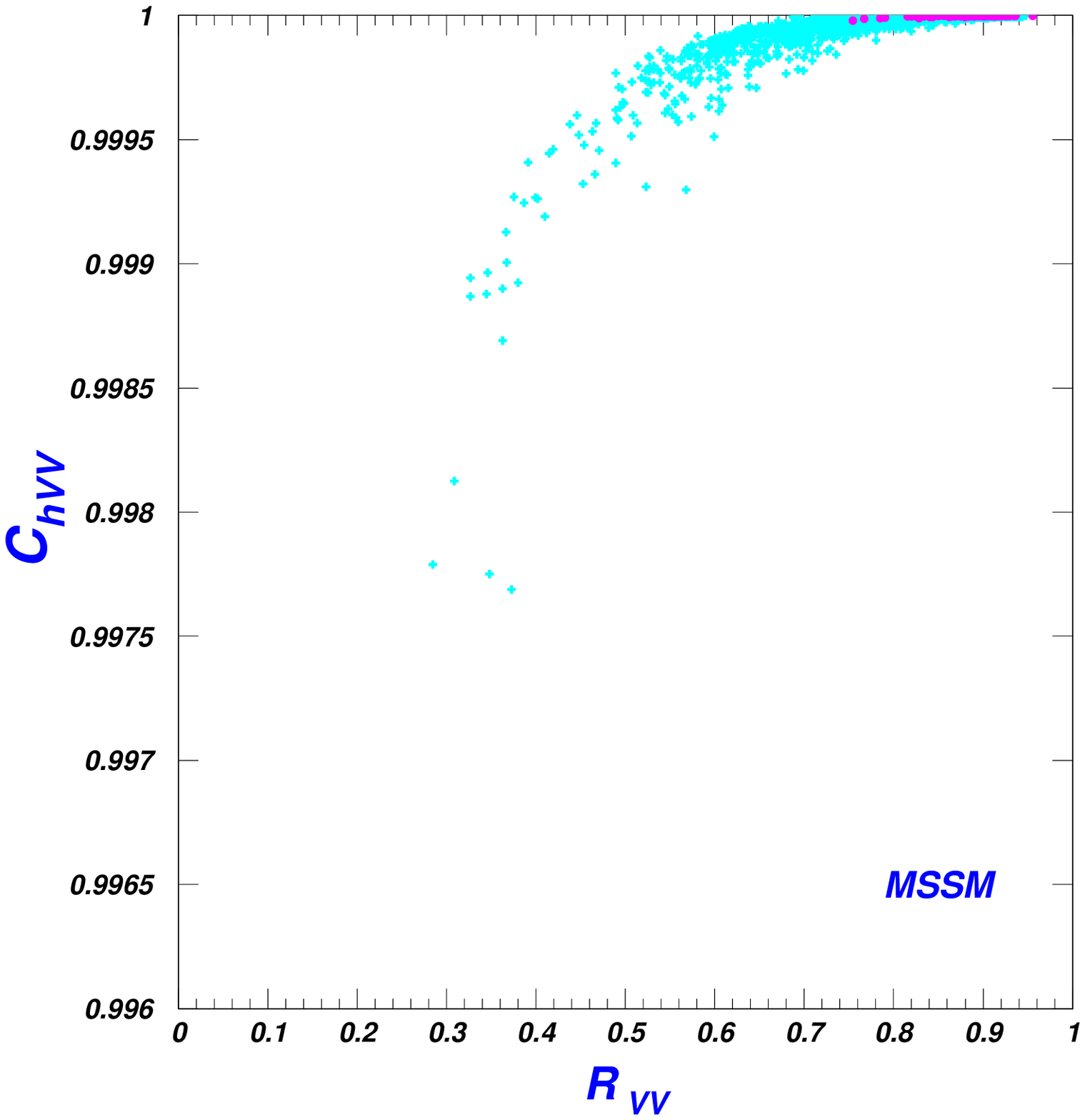}\hspace{0.1cm}
\includegraphics[width=7cm]{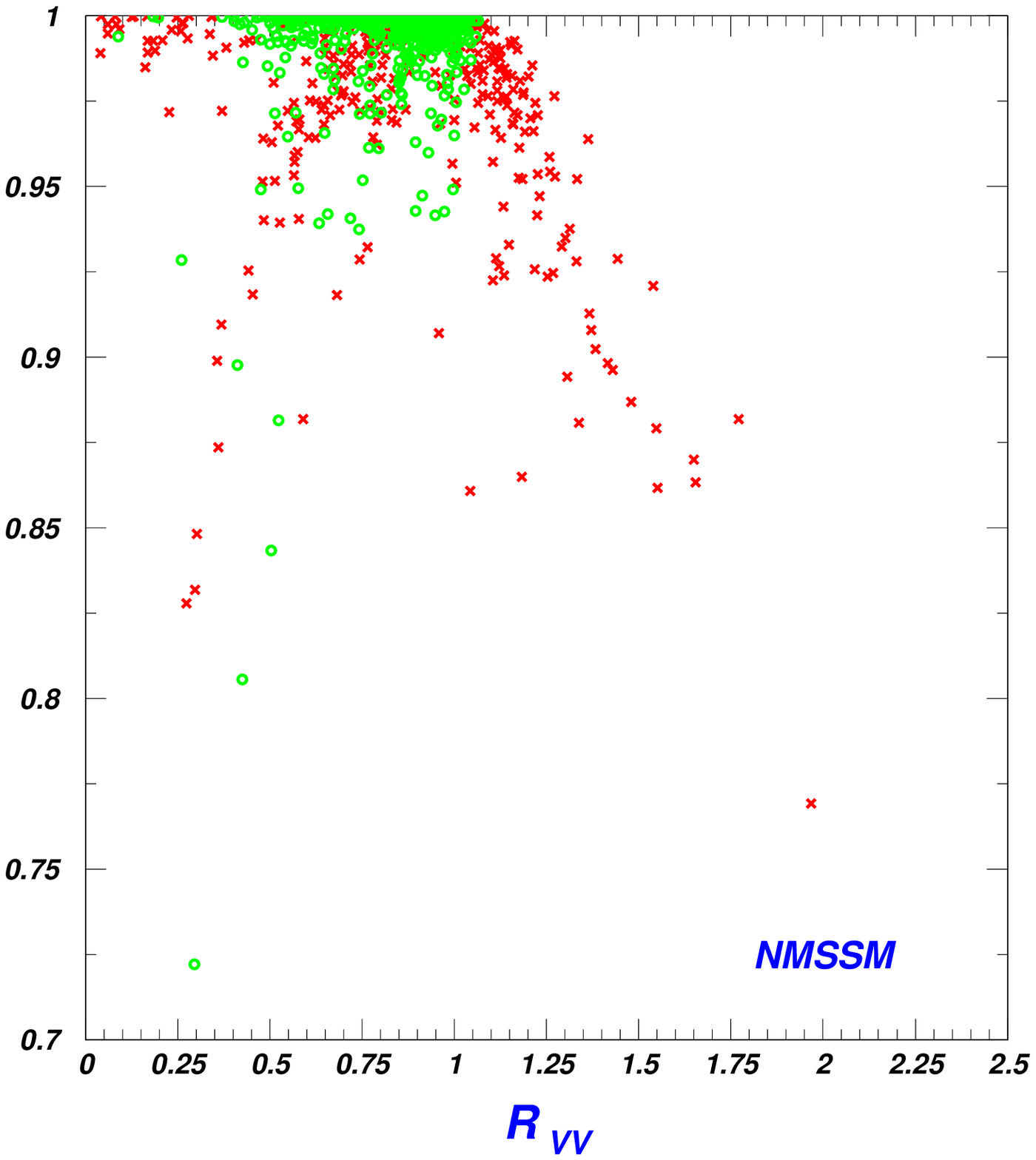}
\vspace*{-0.5cm}
\caption{Same as Fig.\ref{fig2}, but showing the signal rate
$R_{VV}\equiv \sigma_{\rm SUSY} ( p p \to h \to VV^\ast)/\sigma_{\rm SM} ( p p \to h \to VV^\ast)$
versus the coupling $C_{hVV}\equiv C^{\rm SUSY}_{hVV}/C^{\rm SM}_{hVV}$. }
\label{fig5}
\end{figure}
Considering the process $ p p \to h \to V V^\ast$ ($V=W,Z$) is another important
Higgs search channel, we in Fig.\ref{fig5} show the signal rate
versus the $hVV$ coupling. This figure shows that in the MSSM, $h$ is highly
SM-like, while in the NMSSM, the singlet component in $h$ may be sizable, especially
in the push-up case, so that $C_{hVV}$ is reduced significantly. The signal
rate $R_{VV}$ also behaves differently in the two models.
In the MSSM, because $C_{hgg} < 1$ and in most cases $C_{hb\bar{b}} > 1$, $R_{VV}$ is
always less than 1 (for $R_{\gamma \gamma} > 1$ it varies between 0.7 and 0.95).
In the NMSSM, however, $R_{VV}$ may exceed 1 and in this case we find
$R_{VV} \simeq R_{\gamma \gamma}$. The reason for such a correlation is the two
quantities have the same origin for their enhancement,
i.e. the suppression of the $h b \bar{b}$ due to the
presence of the singlet component in $h$.

\begin{figure}[t]
\includegraphics[width=6.5cm]{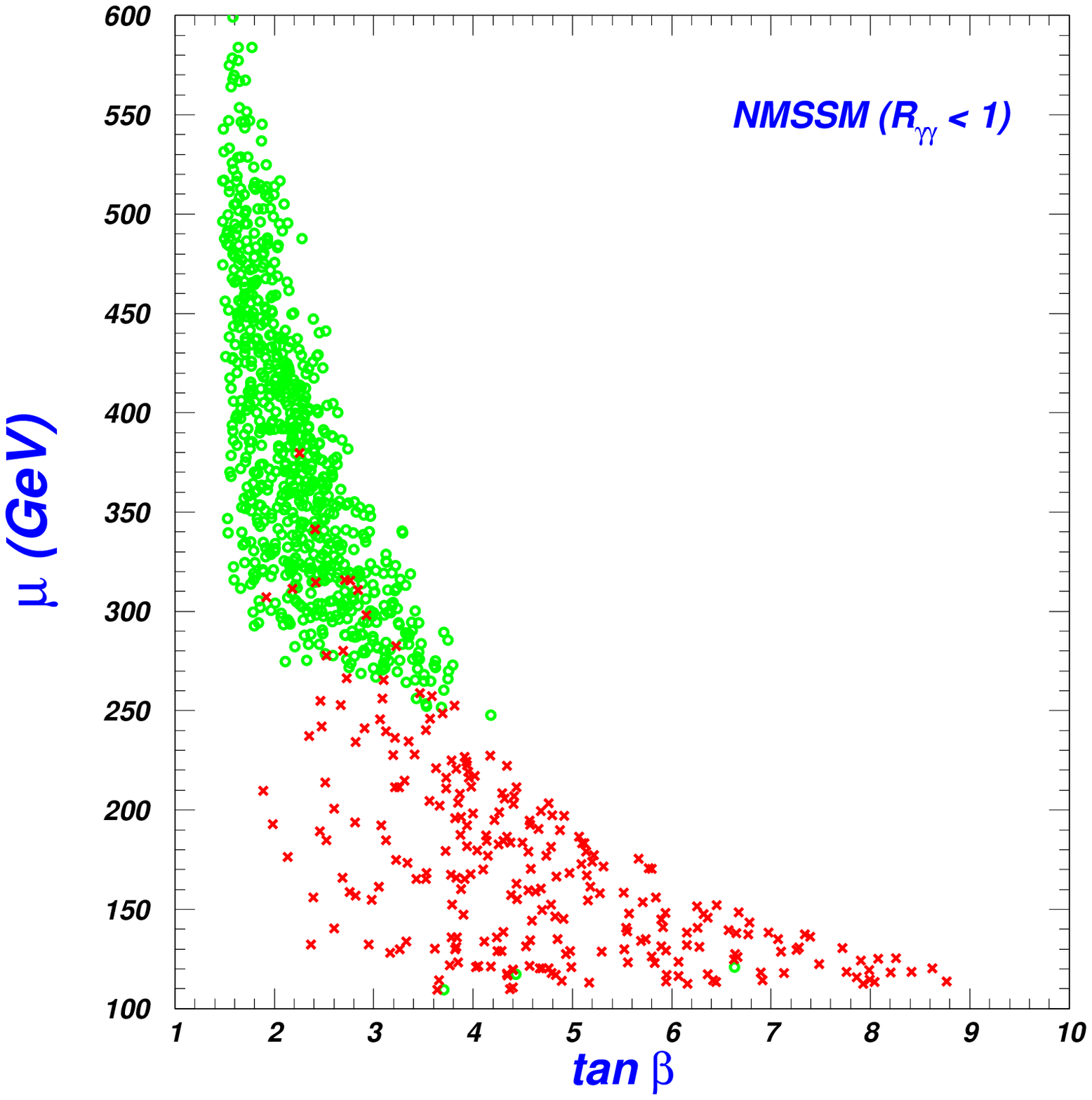}\hspace{0.2cm}
\includegraphics[width=6.5cm]{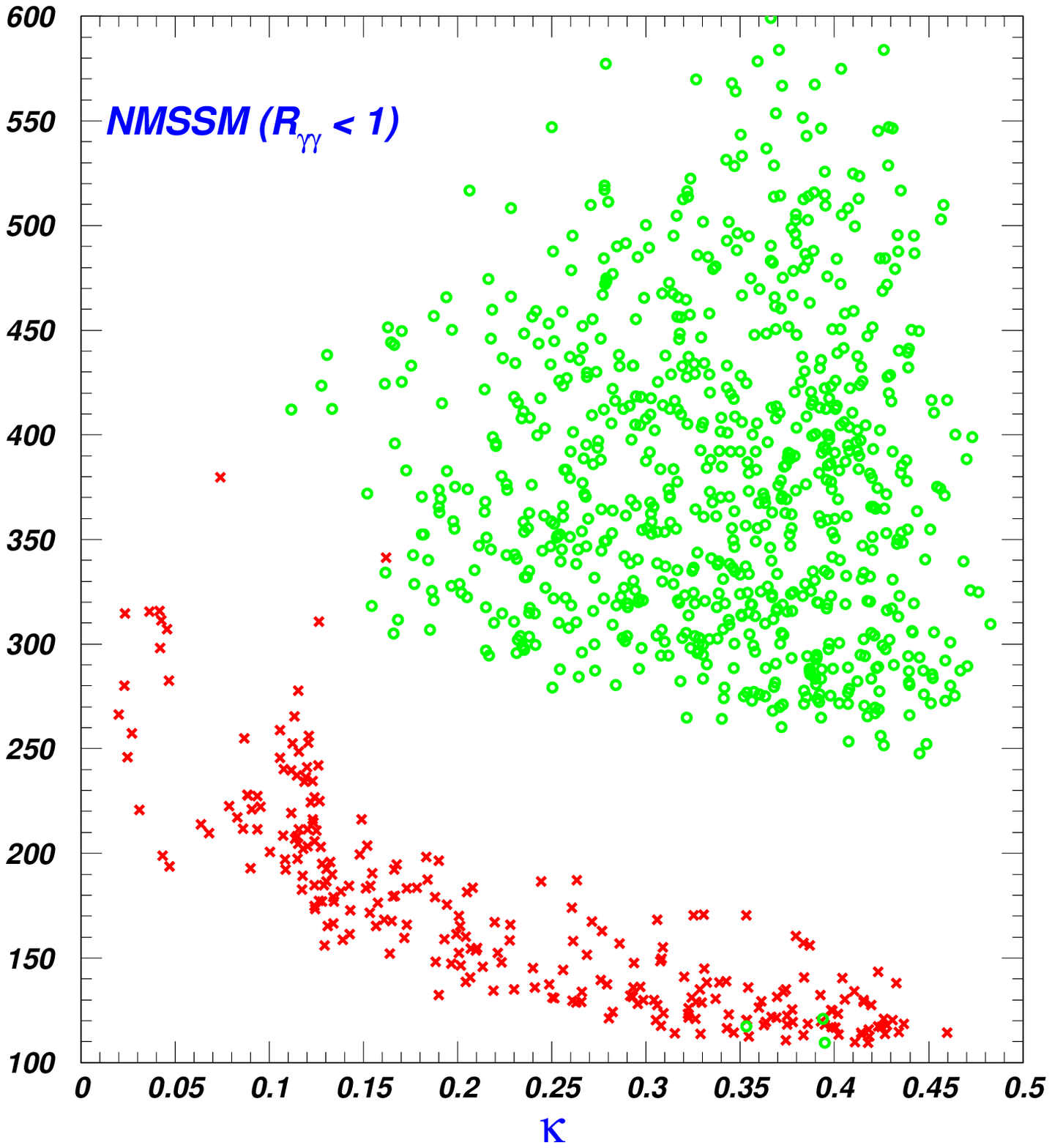}
    \includegraphics[width=6.5cm]{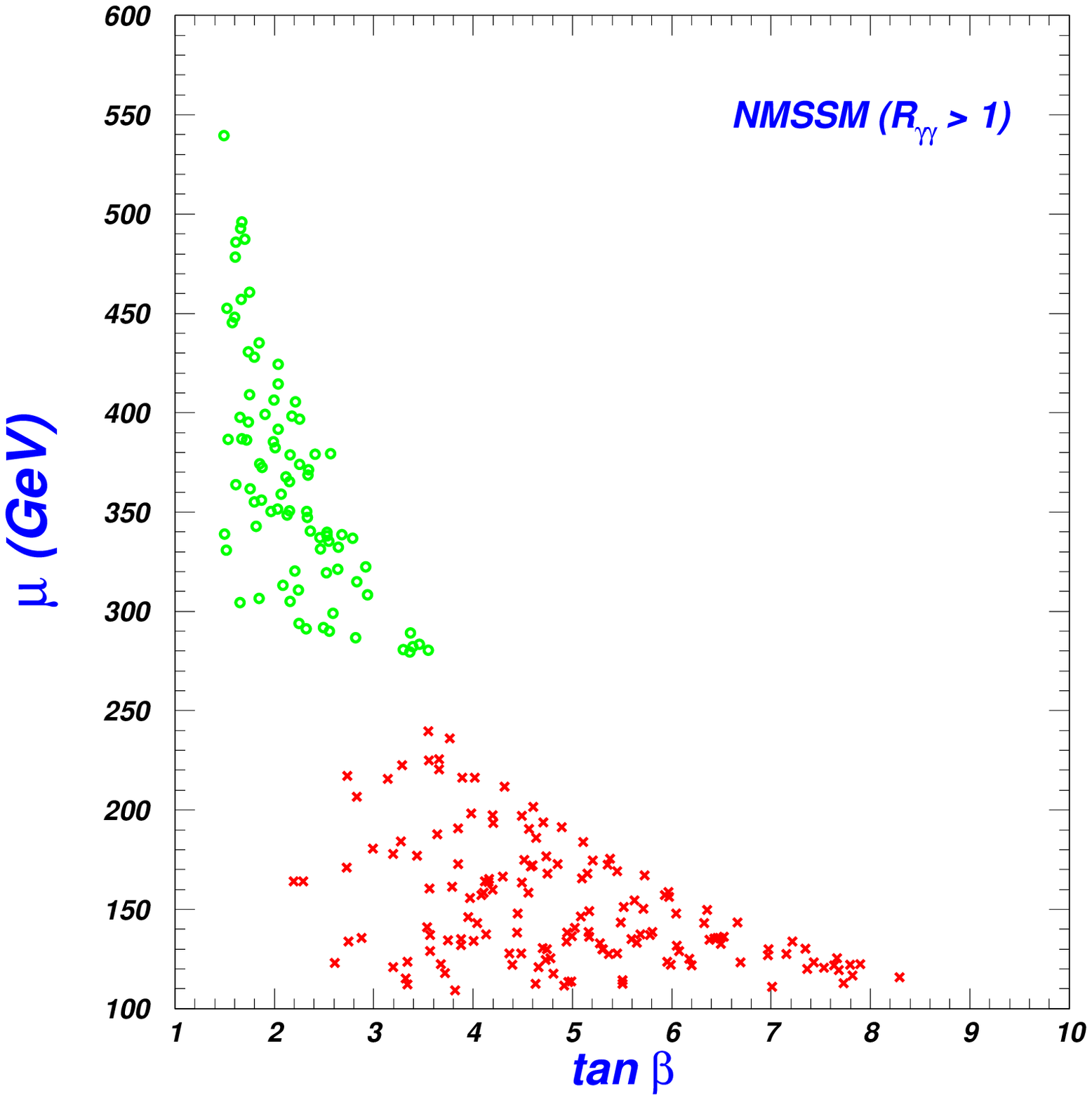}\hspace{0.2cm}
    \includegraphics[width=6.5cm]{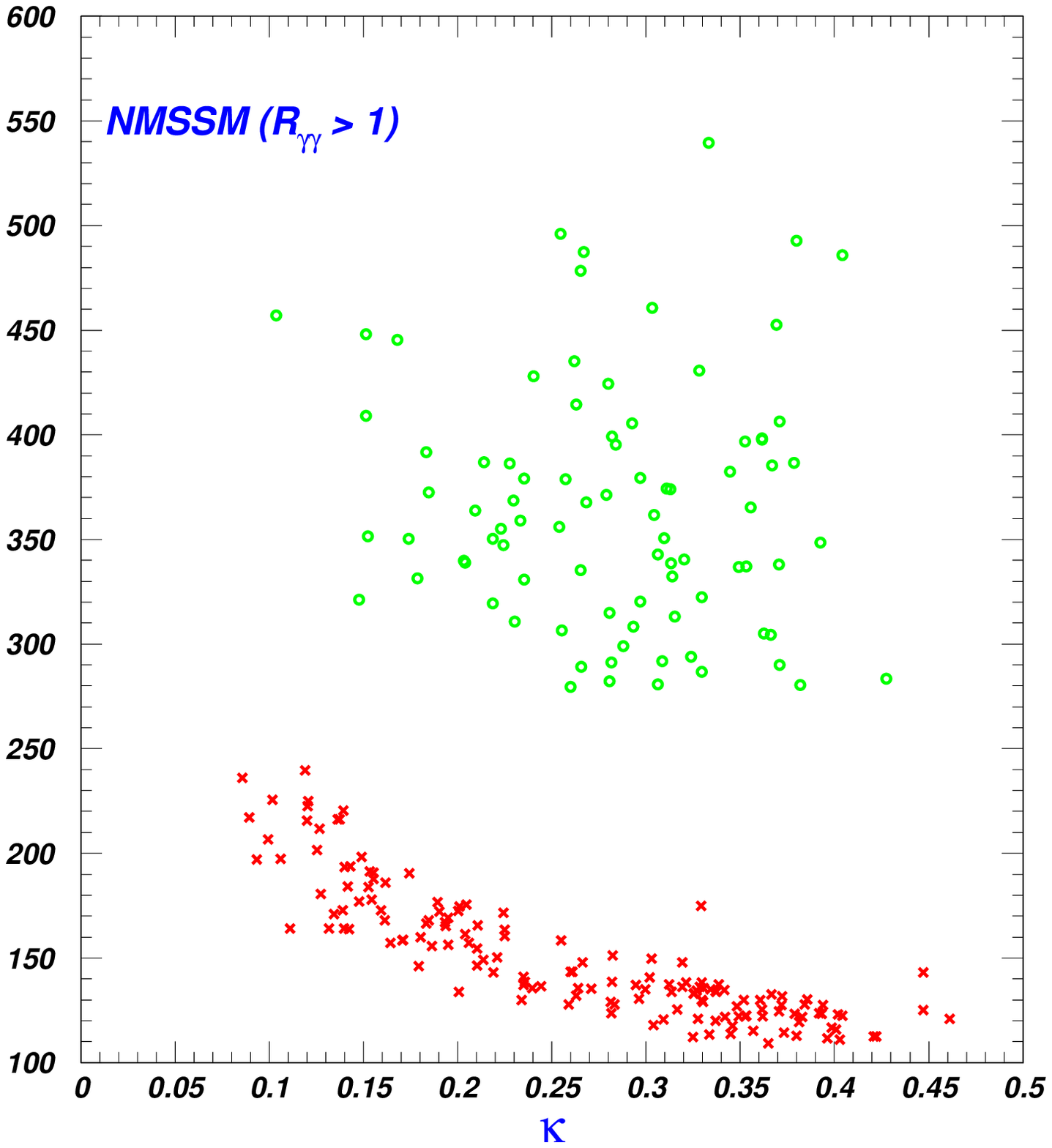}
    \caption{Same as Fig.2, but projected
    in the planes of $\mu$ versus $\tan\beta$ and versus $\kappa$ respectively. }
\label{fig6}
\end{figure}
Next we investigate the favored parameter space of the NMSSM to predict $m_h \simeq 125 {\rm GeV}$.
As introduced in Sec. II,  besides the soft parameters in the stop sector, the sensitive parameters
include $\tan \beta$, $\mu$, $\kappa$ as well as $M_A$.
In Fig.\ref{fig6}, we project the surviving samples
in the planes of $\mu$ versus $\tan\beta$ and versus $\kappa$.
This figure shows three distinctive characters
for the allowed parameters. The first is that $\tan \beta$ must be moderate,
below 4 and 9 for the pull-down and the push-up case, respectively.
Two reasons can account for it. One is that in the NMSSM with large $\lambda$,
the precision electroweak data, i.e. the constraint (4),
strongly disfavor a large $\tan \beta$ \cite{cao08}.
The other reason is that, as far as $\lambda > 0.53$ is concerned,
a moderate $\tan \beta$ is welcomed to enhance the tree
level value of $m_h^2$ (i.e. ${\cal{M}}^2_{22}$) so that even without
heavy stops, $m_h$ can still reach $125 {\rm GeV}$.
Moreover, since the ($S_2,S_3$) mixing is to
reduce the value of $\tilde{M}^2_{11}$ in Eq.(\ref{Re-M}) in the pull-down case,
a larger $\tilde{M}^2_{22}$ (or equivalently a smaller $\tan \beta$) is favored by
the Higgs mass. The second character is that
$\kappa \mu$ in the push-up case is usually much smaller than that in the pull-down
case.
This is because, as we introduced in Sec. II, a large $\kappa \mu$ is needed by
the pull-down case to enhance $\tilde{M}^2_{22}$ in Eq.(\ref{Re-M}).
The third character is obtained by comparing the parameter regions in the upper
panels with those in the lower panels, which shows that $R_{\gamma \gamma} > 1$
puts a lower bound on $\kappa$, i.e. $\kappa \gtrsim 0.1$.
The underlying reason is that for $\kappa < 0.1$, the dark matter will be light
and singlino-like, and to get its currently measured relic density,
the dark matter must annihilate in the early universe by exchanging a light
Higgs boson \cite{cao-xnmssm}. In this case, $h$ mainly decays into the light bosons,
which in return will suppress the di-photon rate.

\begin{figure}[t]
\centering
\includegraphics[width=7cm]{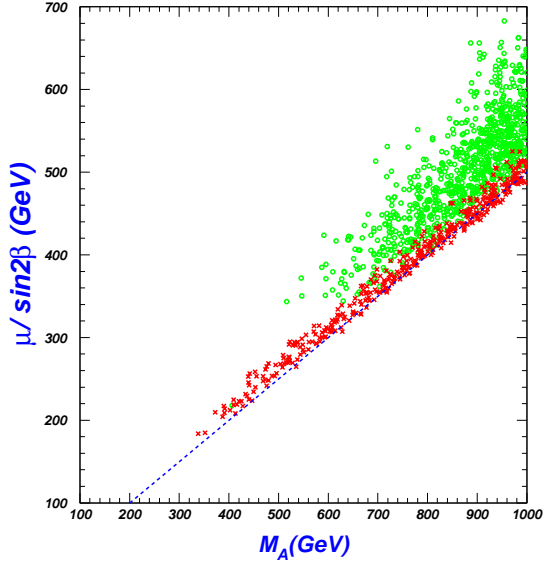}
\vspace*{-0.5cm}
\caption{Same as Fig.\ref{fig6}, but showing the correlation between $M_A$ and $\mu/\sin 2 \beta$.
The dashed line denotes the relation $M_A \sin 2\beta/\mu = 2$.}
\label{fig7}
\end{figure}
In Fig.\ref{fig7} we show the correlation of $M_A$ with $\mu/\sin 2 \beta$.
This figure indicates that $M_A \gtrsim 300 {\rm GeV}$ for the push-up case
and $M_A \gtrsim 500 {\rm GeV}$ for the pull-down case, which is in agreement
with our expectation.  In fact, we checked each surviving sample and found it
satisfies the condition: $M_A^2 \gg {\cal M}^2_{22} \gg {\cal M}^2_{12}$ and
$(M_A^2 - {\cal M}^2_{33}) \gg {\cal M}^2_{13}$, so the samples can be well
described by scenario II. We also checked that the mixing of the
field $S_1$ with $S_2$/$S_3$ is small and $M_A$ is approximately
the heaviest CP-even Higgs boson mass. Fig.\ref{fig7} also indicates that
the relation $M_A \sin 2\beta/\mu = 2 $ is maintained quite well in the push-up
case, but is moderately spoiled in the pull-down case. The reason is,
as we introduced in Sec. II, the requirement that ${\cal{M}}^2_{23}$ should be
moderately small actually implies $C_A \sim 0 $ with $C_A = 1 - (\frac{M_A \sin 2\beta}{2 \mu} )^2
-\frac{\kappa}{2 \lambda}\sin2\beta $. In the push-up case
the third term in $C_A$ is not important, while in the pull case,  although it is several
times smaller than the second term, it may not be negligible. We checked our results and
found $|C_A| \lesssim 0.2 $ and $ 1.4 \lesssim M_A \sin 2\beta/\mu \lesssim 2 $ for all
the surviving samples.

About the NMSSM with $\lambda > 0.53$, three points should be noted. The first is, from
the results presented in Fig.\ref{fig6}, one may find the presence of a smuon and/or
a chargino lighter than $250 {\rm GeV}$. This is because the surviving samples  are
characterized either by $\tan \beta < 4$ or $\mu < 250 {\rm GeV}$
or by both in the NMSSM (see Fig.\ref{fig6}).
Then to explain the discrepancy of muon anomalous
magnetic moment $m_{\tilde{\mu}} \leq 250 {\rm GeV}$ is needed
for a low $\tan\beta$, and $m_{\tilde{\chi}^\pm} \simeq \mu $ implies
$m_{\tilde{\chi}^\pm} \leq 250 {\rm GeV}$. We numerically
checked the validity of this conclusion. The second point is, although
$M_A \sin 2\beta/\mu \simeq 2 $ may be regarded as a new source of fine tuning
in the NMSSM, it is rather predictive to get the value of $M_A$ once
$\mu$ and $\tan \beta$ are experimental determined. Finally, we note the favored region
for $\mu$ and $\tan \beta$ shown in Fig.\ref{fig6} does not overlap with
that in Fig.\ref{fig4}. This may be used to discriminate the models.

\begin{figure}[t]
\centering
\includegraphics[width=7cm]{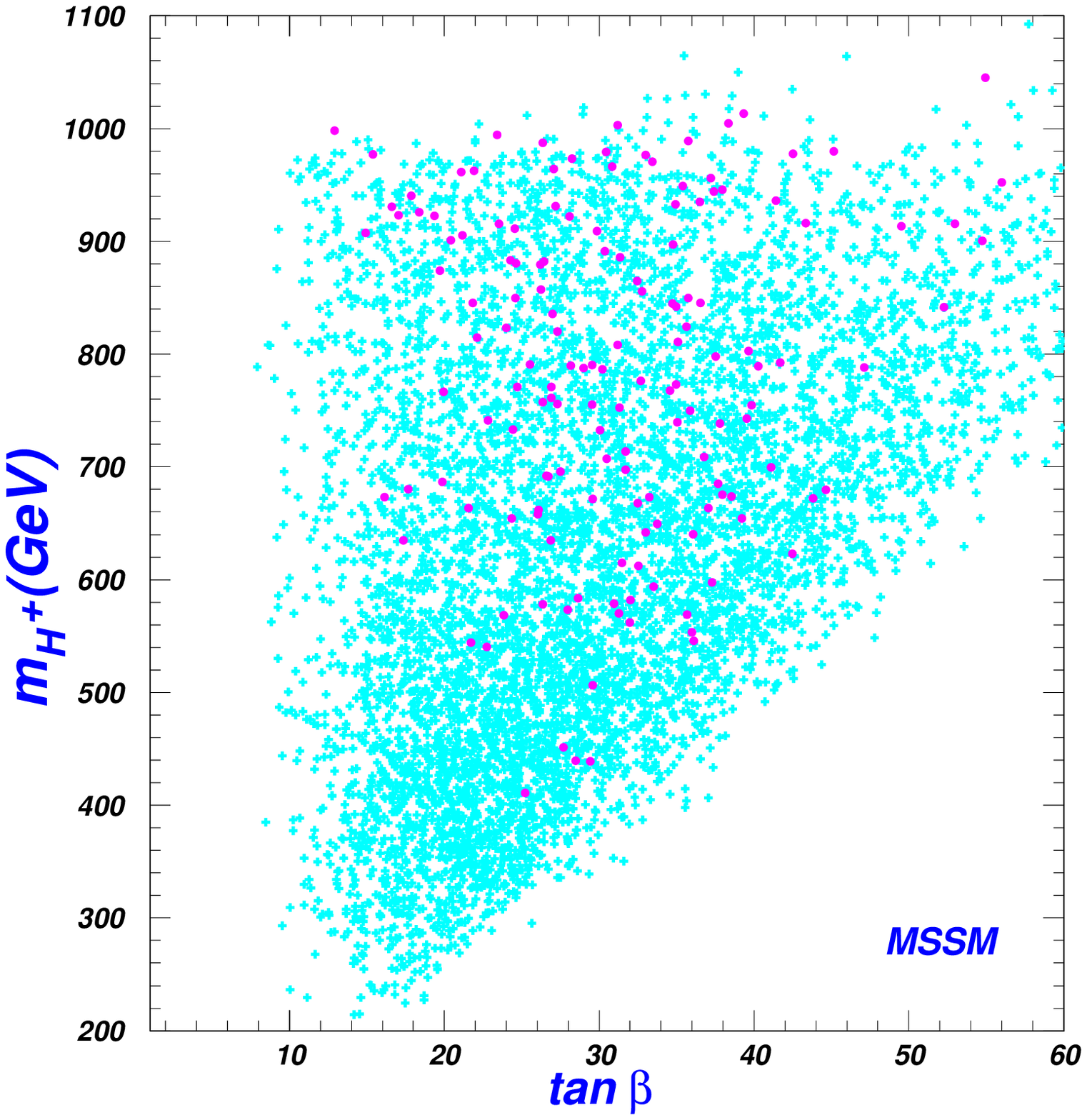}\hspace{0.1cm}
\includegraphics[width=7cm]{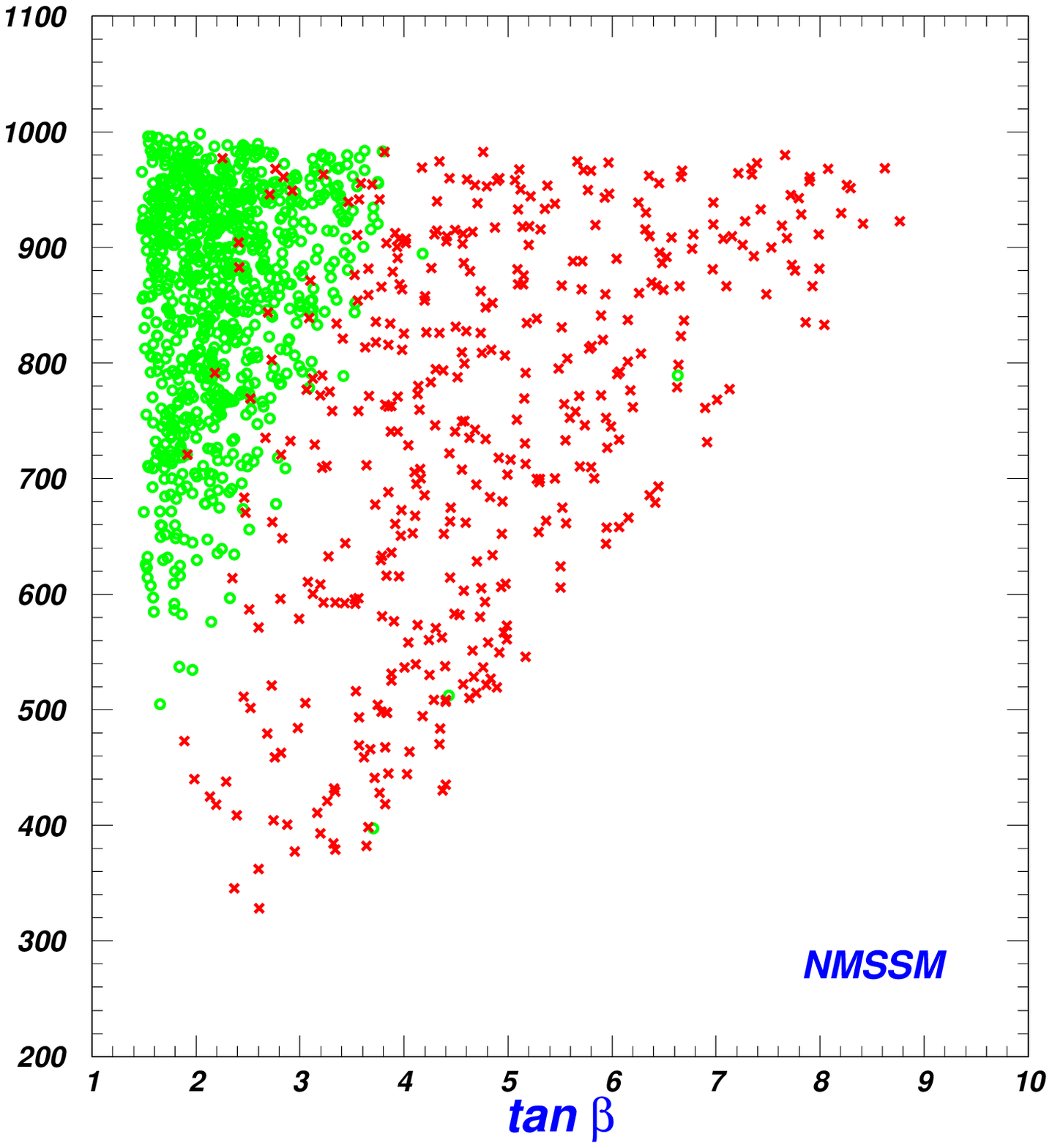}
\vspace*{-0.5cm}
\caption{Same as Fig.\ref{fig5}, but projected on the plane of the charged
Higgs boson mass and $\tan\beta$.}
\label{fig8}
\end{figure}
Finally, we briefly describe other implications of $m_h \simeq 125 {\rm GeV}$
in the SUSY models. In Fig.\ref{fig8} we project the surviving samples on the
plane of $\tan \beta$ versus $m_{H^+}$ with $H^+$ denoting the charged Higgs boson.
This figure shows that $H^+$ must be heavier than about $200 {\rm GeV}$ in the MSSM.
This bound is much higher than the corresponding LEP bound, which is about $80 {\rm GeV}$.
For the NMSSM with a large $\lambda$, the bound can be further pushed up to about
$300 {\rm GeV}$.
This figure also indicates that in the MSSM, $\tan \beta$ may reach 35 for
$m_{H^+}=400 {\rm GeV}$. Then based on the MC simulation by the ATLAS
collaboration \cite{ATLAS0901}, one may expect that the charged Higgs may be
observable from the process $p p \to t H^- \to b W \tau \nu_\tau$ at the early
stage of the LHC.  However, this may be impossible. The reason is,
for relatively light $H^+$ and large $\tan \beta$, $\mu$ must be large to
satisfy the constraint from dark matter direct detection experiments such
as XENON100. This will greatly suppress the $\bar{t}bH^+$ coupling \cite{Cao:2010ph}.
For the NMSSM, the hope to observe $H^+$ is also dim because $\tan \beta$ is small.

\begin{figure}[t]
\centering
\includegraphics[width=7cm]{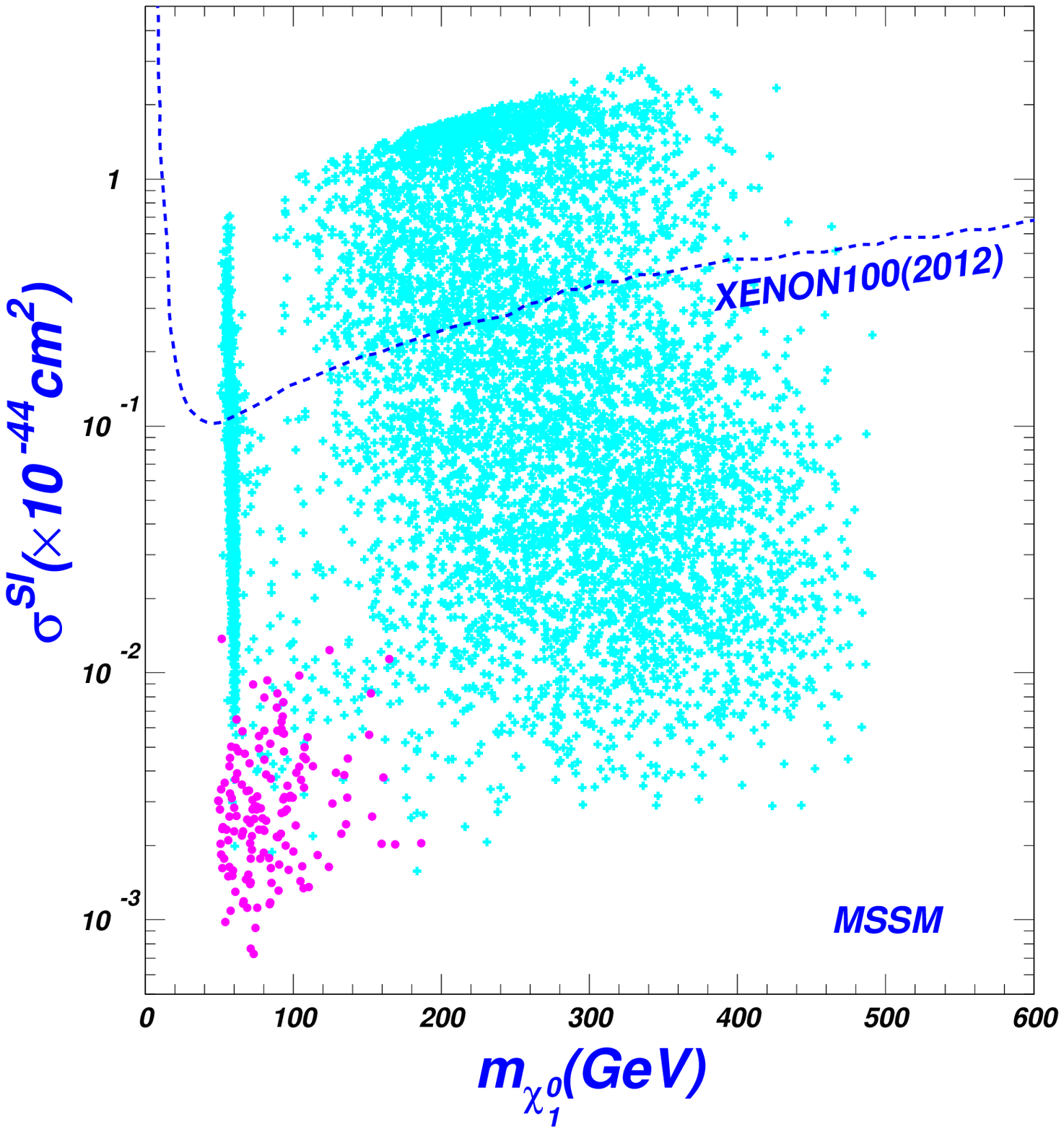}\hspace{0.1cm}
\includegraphics[width=7cm]{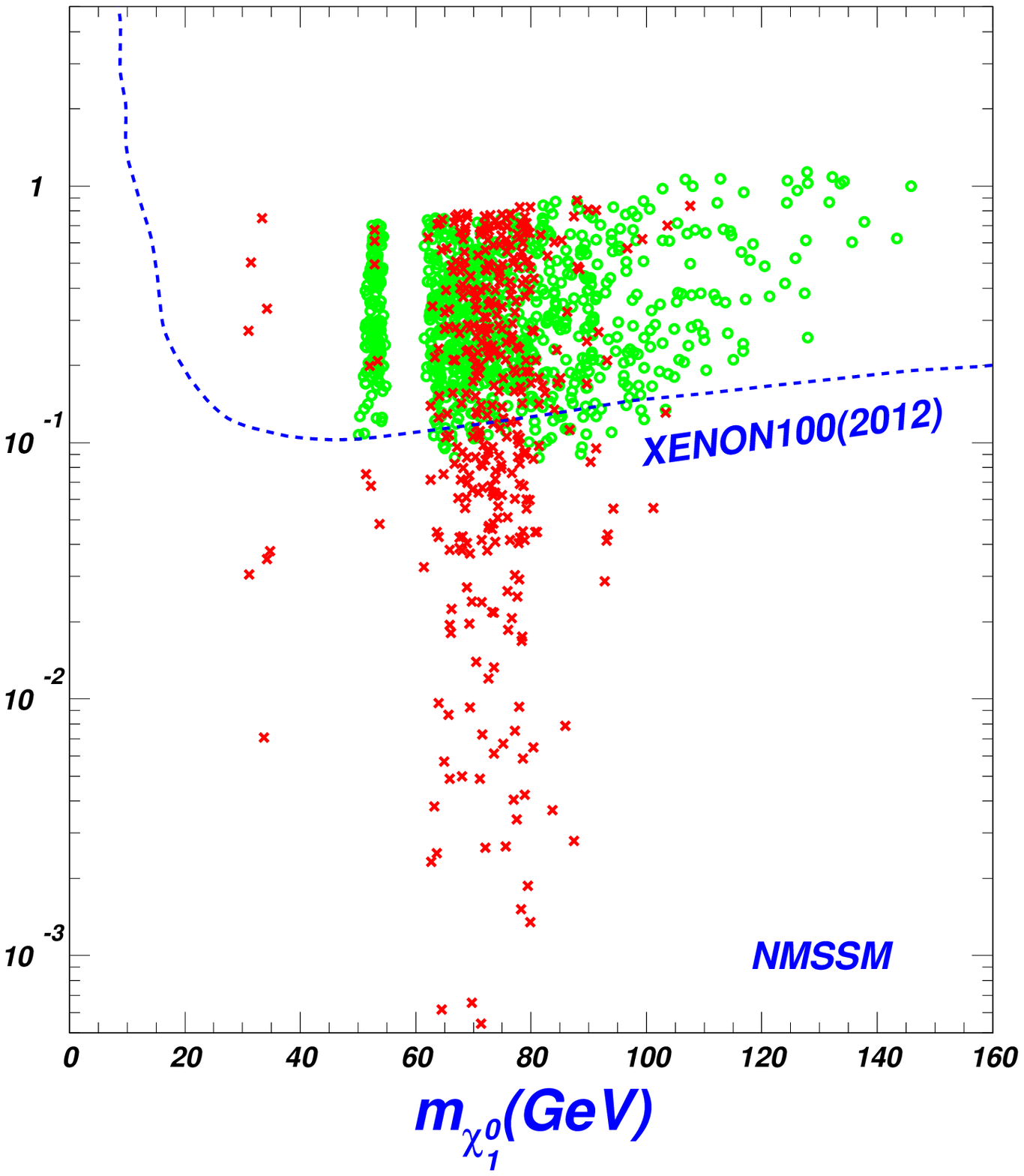}
\vspace*{-0.5cm}
\caption{Same as Fig.\ref{fig2}, but exhibiting
the spin-independent $\chi$-nucleon scattering cross section
 as a function of the dark matter mass.}
\label{fig9}
\end{figure}
In Fig.\ref{fig9} we show the spin-independent elastic scattering between dark
matter and nucleon. We use the formula presented in the Appendix of \cite{cao-dark}
to calculate the scattering rate. As expected, the XENON100 (2012) data
to be released in near future will further exclude some samples,
especially the pull-down case of the NMSSM will be strongly disfavored
if XENON100 (2012) fails to find any evidence of dark matter (assuming the
grand unification relation of the gaugino mass).
From the left of Fig.\ref{fig9} one can 
learn that for the samples with $R_{\gamma \gamma} > 1$
in the MSSM, the scattering rate is small, usually at least one order below than
the sensitivity of the XENON100 (2012).

\section{Conclusion}
Motivated by the recent LHC hints of a Higgs boson around 125 GeV,
we assume a SM-like Higgs with the mass 123-127 GeV and study its
implication in low energy SUSY by comparing the MSSM and NMSSM.
Under various experimental constraints at $2\sigma$ level
(including the muon $g-2$ and the dark matter relic density), we
scanned over the parameter space of each model.
Then in the parameter space allowed by current experimental constraints
and also predicting a SM-like Higgs in 123-127 GeV, we examined the properties
of the sensitive parameters and calculated the rates of the di-photon signal
and the $VV^*$ ($V=W,Z$) signals at the LHC. Among our various findings the typical
ones are: (i)  In the MSSM the top squark and $A_t$ must be large and thus incur some fine-tuning,
which can be much ameliorated in the NMSSM; (ii) In the MSSM a light $\tilde{\tau}$
is needed to make the di-photon rate of the SM-like Higgs exceed its SM
prediction, while the NMSSM has more ways in doing this;
(iii) In the MSSM the signal rates of $pp\to h\to VV^*$ at the LHC are 
never enhanced compared with their SM predictions, 
while in the NMSSM they may be enhanced; (iv) A large part
of the parameter space so far survived will be soon covered by the expected
XENON100(2012) sensitivity (especially for the NMSSM).

Therefore, although the low energy SUSY can in general accommodate a SM-like Higgs boson near
125 GeV and enhance its di-photon signal rate at the LHC, not all models of low energy SUSY
are equally competent if they are required to satisfy all current experimental constraints.
From our present study and some other studies in the literature, we conclude:
\begin{itemize}
\item The fancy CMSSM/mSUGRA is hard to give a 125 GeV SM-like Higgs boson \cite{125Higgs}.
\item The MSSM
can give such a 125 GeV Higgs and can also enhance its di-photon signal rate at the LHC, which,
however, will incur some fine-tuning.
\item The nMSSM (the nearly minimal SUSY model) can give a 125
GeV SM-like Higgs, but severely suppress its di-photon signal rate at the LHC \cite{diphoton3}.
\item The NMSSM is so far the best model to accommodate such a 125 GeV Higgs; it can naturally
(without fine-tuning) predict such a SM-like Higgs mass and readily enhance its di-photon signal
rate at the LHC. At the same time, in a large part of its parameter space, this model can also
enhance the signal rates $pp\to h\to VV^*$ ($V=Z,W$) at the LHC and predict a large scattering
rate of dark matter and nucleon at the XENON100. So the interplay of LHC and XENON100 will soon
allow for a good test of this model.
\end{itemize}

\section*{Acknowledgement}
This work was supported in part by the National Natural
Science Foundation of China (NNSFC) under grant Nos. 10821504,
11135003, 10775039, 11075045, by Specialized Research Fund for
the Doctoral Program of Higher Education with grant No. 20104104110001, and  by the Project of Knowledge
Innovation Program (PKIP) of Chinese Academy of Sciences under grant
No. KJCX2.YW.W10.

\end{document}